\documentclass[12pt,journal,draftcls,letterpaper,onecolumn]{IEEEtran}
\usepackage{microtype,times}
\usepackage[cmex10]{amsmath}
\usepackage{cite,array,eqparbox}

\usepackage[overload]{textcase}
\usepackage{anyfontsize}
\usepackage{amsmath,amssymb,amsthm}
\usepackage{tikz}
\usetikzlibrary{trees}
\usepackage{rotating}
\usepackage{graphicx,graphics,epsfig,float,verbatim,array,bm,multirow,color}
\usepackage{amsthm}
\usepackage[left=1.5cm,top=1.5cm,bottom=1.5cm,right=1.5cm]{geometry}
\newcommand{\R}{{\mathbb R}}

\newcommand {\sr}{\stackrel}

\newcommand {\eqv}{\equiv}
\newcommand {\Real}{\mathbb{R}}

\newcommand {\td}{\tilde}

\newcommand {\ra}{\rightarrow}
\newcommand {\Ra}{\Rightarrow}

\newcommand {\La}{\Leftarrow}

\newcommand {\ral}{\longrightarrow}

\newcommand {\ub}{\underbrace}

\newcommand {\vep}{\varepsilon}

\newcommand {\del}{\partial}

\newcommand {\ld}{\lambda}

\newcommand {\al}{\alpha}

\newcommand {\bfr}{\begin{flushright}}
\newcommand {\efr}{\end{flushright}}
\newcommand {\bfl}{\begin{flushleft}}
\newcommand {\efl}{\end{flushleft}}
\newcommand {\nn} {\nonumber}

\newcommand {\txt}{\textrm}
\newcommand {\bd}{\begin{document}}
\newcommand {\ed}{\end{document}}

\newcommand {\be}{\begin{equation}}
\newcommand {\ee}{\end{equation}}
\newcommand {\bea}{\begin{eqnarray}}
\newcommand {\eea}{\end{eqnarray}}
\newcommand {\ba}{\begin{aligned}}
\newcommand {\ea}{\end{aligned}}
\newcommand {\bit}{\begin{itemize}}
\newcommand {\eit}{\end{itemize}}
\newcommand {\ul}{\underline}
\newcommand {\txtc}{\textcolor}

\newcommand {\ad}{\textrm{ad}}
\newcommand {\Ad}{\textrm{Ad}}
\def\A{{\cal A}}\def\B{{\cal B}}\def\C{{\cal C}}\def\D{{\cal D}}\def\E{{\cal E}}\def\F{{\cal F}}\def\G{{\cal G}}\def\H{{\cal H}}\def\I{{\cal I}}
\def\J{{\cal J}}\def\K{{\cal K}}\def\L{{\cal L}}\def\M{{\cal M}}\def\N{{\cal N}}\def\O{{\cal O}}\def\P{{\cal P}}\def\Q{{\cal Q}}\def\R{{\cal R}}
\def\S{{\cal S}}\def\T{{\cal T}}\def\U{{\cal U}}\def\V{{\cal V}}\def\W{{\cal W}}\def\X{{\cal X}}\def\Y{{\cal Y}}\def\Z{{\cal Z}}

\def\xbar{\bar{x}}\def\ybar{\bar{y}}\def\zbar{\bar{z}}\def\kbar{\bar{k}}\def\pbar{\bar{p}}

\newtheorem{thm}{Theorem}[section]
\newtheorem{lmm}{Lemma}[section]
\newtheorem{dfn}{Definition}[section]
\newtheorem{crl}{Corrollary}[section]
\newtheorem{prp}{Proposition}[section]
\newtheorem{example}{Example}[section]
\newtheorem{prb}{Problem}[section]
\newtheorem{alg}{Algorithm}[section]
\newtheorem{rmk}{Remark}[section]
\theoremstyle{definition}\newtheorem*{rmks}{Remarks}
\newcommand*{\Cdot}[1][1.25]{%
  \mathpalette{\CdotAux{#1}}\cdot%
}
\newdimen\CdotAxis
\newcommand*{\CdotAux}[3]{%
  {%
    \settoheight\CdotAxis{$#2\vcenter{}$}%
    \sbox0{%
      \raisebox\CdotAxis{%
        \scalebox{#1}{%
          \raisebox{-\CdotAxis}{%
            $\mathsurround=0pt #2#3$%
          }%
        }%
      }%
    }%
    \dp0=0pt %
    \sbox2{$#2\bullet$}%
    \ifdim\ht2<\ht0 %
      \ht0=\ht2 %
    \fi
    \sbox2{$\mathsurround=0pt #2#3$}%
    \hbox to \wd2{\hss\usebox{0}\hss}%
  }%
}
\begin{document}
%\title{Convex Objective-Based Fusion in Distributed Detection: Architecture and Performance Analysis of Interactive Fusion}
\title{Interactive Distributed Detection: Architecture and Performance Analysis}
\author{Earnest~Akofor,~and~Biao~Chen,~\IEEEmembership{Senior Member, IEEE}
\thanks{E. Akofor and B. Chen are with the Department
of Electrical Engineering and Computer Science, Syracuse University, Syracuse,
NY, 13244 USA e-mail: $\{$eakofor,bichen$\}$@syr.edu.}%
}

%\markboth{Journal of \LaTeX\ Class Files,~Vol.~1, No.~8,~August~2002}{Shell \MakeLowercase{\textit{et al.}}: Bare Demo of IEEEtran.cls for Journals}

% If you want to put a publisher's ID mark on the page
% (can leave text blank if you just want to see how the
% text height on the first page will be reduced by IEEE)
%\pubid{0000--0000/00\$00.00~\copyright~2002 IEEE}

%\specialpapernotice{(Invited Paper)}

\maketitle
%\tableofcontents

\begin{abstract}
 This paper studies the impact of interactive fusion on detection performance in tandem fusion networks with conditionally independent observations. Within the Neyman-Pearson framework, two distinct regimes are considered: the fixed sample size test and the large sample test. For the former, it is established that interactive distributed detection may strictly outperform the one-way tandem fusion structure. However, for the large sample regime, it is shown that interactive fusion has no improvement on the asymptotic performance characterized by the Kullback-Leibler (KL) distance compared with the simple one-way tandem fusion. The results are then extended to interactive fusion systems where the fusion center and the sensor may undergo multiple steps of memoryless interactions or that involve multiple peripheral sensors, as well as to interactive fusion with soft sensor outputs.
\end{abstract}


\let\thefootnote\relax\footnotetext{This work was supported by Air Force Office of Scientific Research under Award FA9550-10-1-0458, by Army Research Office under Award W911NF-12-1-0383, and by National Science Foundation under Award CCF1218289. The material in this paper was presented in part at the 2013 IEEE International Conference on Acoustic, Speech, and Signal Processing, Vancouver, Canada, May 2013.}

\section{Introduction}
% \PARstart{A}{demo} file is ....
% \PARstart{A}{}demo file is ....
% \PARstart{T}{his demo} file is ....
\PARstart{A}{} simple tandem sensor network typically consists of two or more sensors, one of them serving as a fusion center (FC) that makes a final decision using its own observation as well as input from the other sensors. Practical constraints often dictate that the input from the other sensors is maximally compressed. The extreme case is that the observation at each one of the other sensors is mapped to a single bit, often referred to as a local decision. Distributed detection with such a tandem network has been relatively well understood under the conditional independence assumption, i.e., the observations at distributed nodes are independent conditioned on a given hypothesis. Specifically, it was known that the optimal local sensor decision rule is in the form of a likelihood ratio test \cite{Varshney:book}. Fusion architecture, and in particular, the impact of communication direction in a two sensor system was studied in \cite{Papastavrow&Athans:92AC,song-et-al,song-et-al2}.

\begin{figure}%[H]
\centering
%\includegraphics[width=9cm,height=5cm]{processes}\\
%\input{fusion-diagrams.pstex_t}% magnification = 50%
%\input{fusion-diagrams2.pstex_t}% magnification = 50%
%\scalebox{2}{\input{fusion-diagrams3.pstex_t}}% magnification = 50%
\scalebox{2}{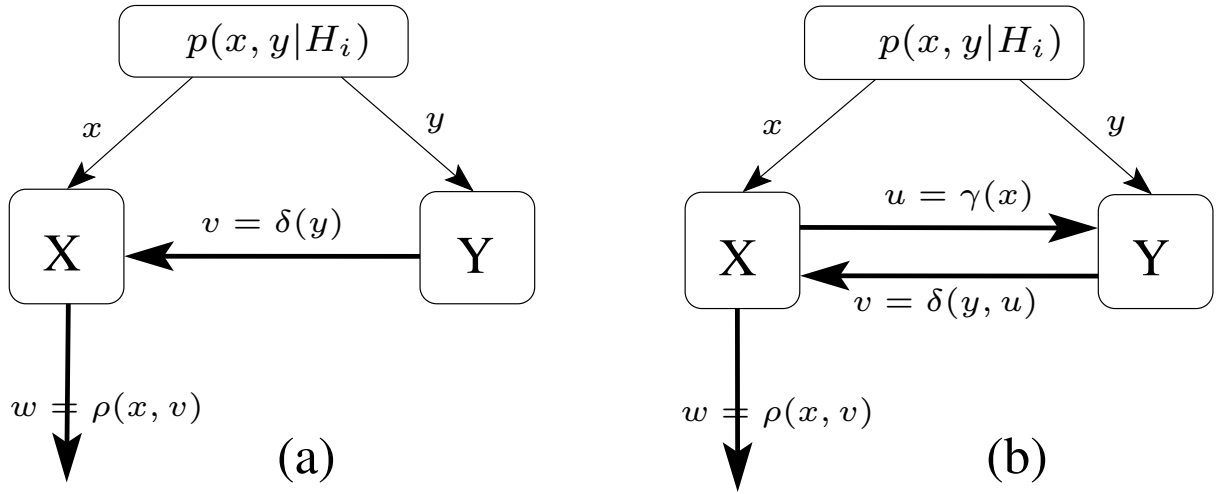}% magnification = 50%
\vspace{-0.2cm}
\caption{(a) One-way tandem fusion (YX process),~~ (b) Interactive fusion (XYX process).}\label{one-and-two-way-diagram}
\vspace{-0.4cm}
\end{figure}

This paper revisits this simple tandem distributed detection network by replacing the static message passing (from the sensor node to the fusion node)  with an interactive one: it is assumed that the FC may send an initial bit to the local sensor based on the observation at the FC. The local sensor then makes a local decision based on its own observation as well as the input from the FC before passing it back to the FC. In the most general setting, as to be discussed in Section \ref{generalize}, multiple rounds of interactions may occur, the interaction may occur between the FC and multiple sensors, and soft (i.e., multi-bit) output may be exchanged. For the most part, we limit ourselves to a single round of interaction between two sensors and we refer to this communication protocol as the so-called interactive fusion. The contrast between the traditional one-way tandem fusion network and the interactive fusion network is illustrated in Fig.~\ref{one-and-two-way-diagram}.

Let $x$ be the observation of sensor X and $y$ be the observation of sensor Y. Then for a one-way tandem network the decision variables $(v,w)$ are based on $x$ and $y$ through dependencies of the form $v=\delta(y),~w=\rho(x,v)$, where $\delta$ and $\rho$ are integer-valued mappings. Similarly, for the interactive model the decision processes based on observations $x$ and $y$ yield outputs $(u,v,w)$, where $u=\gamma(x),~v=\delta(y,u),~w=\rho(x,v)$, with integer-valued mappings $\gamma$, $\delta$, and $\rho$. For simplicity, we refer to the fusion architecture in Fig.~\ref{one-and-two-way-diagram}(a) as the YX process whereas to that in Fig.~\ref{one-and-two-way-diagram}(b) as the XYX process. We assume, throughout this paper, that the observations $x$ and $y$ are conditionally independent given any hypothesis under test.

This interactive fusion network has been studied under the Bayesian framework and  was shown to improve the error probability performance for fixed-sample size test \cite{s-a-c}. A similar model was also used in \cite{Xiang-Kim:Allerton12} to study the test of independence where the interaction is subject to a rate constraint. This interactive fusion model is different from the traditional feedback setting considered in \cite{alhakeem-varshney,Tay&Tsitsiklis,Shalaby&Papamarcou} where a global FC first receives information from local sensors and sends a summary information back to the local sensors.

This paper focuses on the Neyman-Pearson (NP) framework and we address whether the additional input from the FC would improve
\begin{enumerate}%[leftmargin=0.5cm]%[leftmargin=*]
\item the performance of the fixed sample size NP test;
\item the asymptotic performance, quantified using the Kullback-Leibler distance which is known to be the error exponent of the type II error for an NP test, a.k.a., the Chernoff-Stein Lemma \cite{cover-thomas}.
\end{enumerate}
We show that while the answer to the first question is affirmative, interactive fusion does not improve the asymptotic performance, i.e., the answer to the second question is negative. We note that in our setup asymptotics is with respect to the number of independent samples taken over time while the number of sensors remains fixed. Our setup is different from \cite{Tay&Tsitsiklis,Shalaby&Papamarcou,Papastavrou-Athans-2,Tay-Tsitsiklis-Win} where asymptotics is taken with respect to the number of sensors.  Preliminary results were reported in \cite{akofor-chen-icassp} where we considered strictly the simple two-sensor system with single bit sensor output. In the present work, detailed proofs and analysis are given along with extensions to more complicated fusion systems as described in Section V.

%Feedback in \cite{alhakeem-varshney} is also from a global FC to local sensors, and is non-interactive.

The analysis for the asymptotic case assumes sample-by-sample processing, i.e., scalar quantization is used at each stage. An alternative approach is to use vector quantization: samples are processed in blocks with block length tending to infinity when warranted. Scalar quantization is chosen as it is much simpler to implement than vector quantization and is much more amenable to analysis. Another important advantage is that it incurs minimum delay as processing incurs zero delay for the first two stages in the interactive scheme. For vector quantization, however, processing at each stage in the interactive fusion scheme is only completed at the end of each block and such delays are cumulative among different stages. We note that scalar quantization has been used before for asymptotic analysis in distributed detection \cite{LLG}.
\begin{comment}
In the asymptotic test, we shall consider sample-by-sample processing. This is because, as pointed out in \cite{LLG}, scalar quantizers are not only simple to implement as compared to vector quantizers, but they are also safer since their relatively simple processing is less likely to introduce large delays capable of degrading overall system performance.
\end{comment}

Our presentation is organized as follows. Section \ref{section_dc} describes the procedure for obtaining decision rules given a convex objective function. The obtained result in Proposition \ref{region_proposition} will be applied in subsequent sections. We demonstrate in Section \ref{section_NP} that interactive fusion does improve performance of a fixed sample size NP test. For the large sample regime, however, we show in Section \ref{section_KL} that interactive fusion does not improve detection performance as characterized using the error exponent. Generalizations of our result to multiple-step memoryless interactive fusion, interactive fusion involving multiple sensors, and with soft sensor outputs are included in Section \ref{generalize}. Section \ref{concl} contains concluding remarks.

{\textbf{Notation:} \it In order to avoid clutter with symbols and obscuring of essential details, we use lower case letters such as $x,y,...$ to denote both random variables and their values. For the same reasons, we use the same symbol $\sum_x$ to denote both summation, when $x$ is a discrete variable, and integration,
when $x$ is a continuous variable. Likewise, $\delta_{xx'}$ denotes the Kronecker delta when $x$ and $x'$ are discrete, and the Dirac delta when $x$ and $x'$ are continuous. Therefore, throughout we apply the familiar functional identity
\bea
{\del f(x)/\del f(x')}=\delta_{xx'},\nn
\eea
where $x$ and $x'$ can be tuples of several discrete or continuous variables, for which we naturally define $\delta_{(x_1,...,x_N)(x'_1,...,x'_N)}$ by
\bea
 \delta_{(x_1,x_2,\cdots
 ,x_N)(x'_1,x'_2,\cdots,x'_N)} ~\triangleq~ \delta_{x_1x'_1} \delta_{x_2x'_2}\cdots \delta_{x_Nx'_N}.\nn
\eea
}

\section{The underlying decision theory}\label{section_dc}
Consider the test of $M$ simple hypotheses
\bea
H_i:~x~\sim~ p_i(x),~~~~i=0,1,...,M-1,
\eea
%due to a processor,
where $x\in\X$ is the observation with distribution
$$p_i(x)~\triangleq~p(x|H_i)$$
under the $i$th hypothesis $H_i$. Under $H_i$, we will denote the probability of a data set $A\subset \X$ by
\bea
P_i(A)=\sum_{x\in A}p_i(x).\nn
\eea
The observation space $\mathcal{X}$ may be of arbitrary dimension. A \emph{decision rule} is a mapping defined as
\bea
\gamma:~x~\mapsto ~i~\in~\{0,1,...,N-1\},
\eea
where $\gamma$ is a deterministic function. We refer to the assignment ~$\gamma(x)=i$~ as a \emph{decision} based on the observation $x$. In this paper, without loss of generality, we assume that the decision output has the same alphabet as the underlying hypothesis; i.e., $N=M$. Thus, $\gamma(x)=i$ can be interpreted as acceptance of the $i$th hypothesis $H_i$.

The desired decision rule $\gamma$ so defined is deterministic in the sense that $p(\gamma(x)=i|x)=\delta_{i,\gamma(x)}$, where $\delta_{a,b}\triangleq 1$ if $a=b$ and $0$ otherwise. Therefore, once $x$ is given $\gamma(x)$ is precisely known. As the optimum decision rule is not necessarily deterministic, we consider the larger set containing all deterministic and nondeterministic decision rules. Let us write the generic decision rule as
\bea
\Gamma:~x~ \mapsto ~i~\in~\{0,1,...,M-1\},
\eea
and let $u=\Gamma(x)$.
Then $\Gamma=\gamma$ denotes a deterministic choice of the decision rule. Recall as in \cite{Tsitsiklis} that the set of $\Gamma$ is the convex hull of the set of $\gamma$. Therefore
\bea
p(\Gamma(x)=i)=\sum_{g}p(g)~p\big(\gamma_g(x)=i\big)
 \eea
where $g$ is a random variable with probability mass, or density, function $p(g)$ and is independent of $x$. The decision process simply picks the appropriate $p(g),$ and hence the desired $p(\Gamma(x)=i)$.

As $u$ is a random variable, making an optimal guess $u=i$ is equivalent to choosing $p(u=i|x)$ such that some \emph{objective function}, which we denote by $S$, is optimized. Here $S$ is a function of $p(u=i|x)$ for all $i=0,1,...,M-1$ and for all data points $x\in\X$.
\begin{comment}
However, since the decision rule is by definition specified per data point, it suffices to separately optimize $S$ with respect to $\{p(u=i|x),~i=0,1,...,M-1\}$ for each data point $x\in \X$.
\end{comment}

In general, $0\leq p(u=i|x)\leq 1,~~i=0,1,...,M-1$. A \emph{deterministic decision rule} is one for which $p(u=i|x)$ takes on only the boundary values $0$ and $1$. For such cases, the decision rule is equivalently expressed as a partition of the data space into disjoint decision regions, i.e.,
\be
\label{optimal1} p(u=i|x)~\triangleq~p(\gamma(x)=i|x)= I_{R_{u=i}}(x),
\ee
where $I_{R_{u=i}}(x)=\delta_{i,\gamma(x)}$ is the indicator function of the region $R_{u=i}=\{x:~\gamma(x)=i\}$, i.e., the \emph{decision region} for the $i$th hypothesis.

In the following proposition, we establish the general structure of the optimal decision rule for an important class of decision problems namely, those with convex objective functions.
\begin{comment}
for which we show in particular that $\gamma(x)$ in (\ref{optimal1}) is given by
\bea
\gamma(x)=\mathop{\arg\max}_i~\del S/\del p_{\txt{opt}}(u=i|x).\nn
\eea
\end{comment}

\begin{prp}\label{region_proposition}
Let $x$ be a random variable or vector, and suppose the objective function $S$ to be maximized is a differentiable convex function of $p(u|x)$ for any given $x$ in the observation sample space $\X$. For each $i$ suppose further that the set of data points
\be
\label{null_set}C_{u=i}=\bigcup_{j\neq i}\left\{x:~{\del S/\del p_{\txt{opt}}(u=i|x)}={\del S/\del p_{\txt{opt}}(u=j|x)}\right\}
\ee
has zero probability measure, where
$${\del S/\del p_{\txt{opt}}(u=j|x)}={\del S/\del p(u=j|x)}|_{p(u=j|x)=p_{\txt{opt}}(u=j|x)}.$$
Then the resulting optimal rule is deterministic, and is given by
\bea
\label{solution}p_{\txt{opt}}(u=i|x)=I_{R_{u=i}}(x),~~~~i=0,1,...,M-1,
\eea
where the $i$th decision region $R_{u=i}$ is specified as
\be
\label{decision-regions}R_{u=i}=\bigcap_{j\neq i}\left\{x:~{\del S/\del p_{\txt{opt}}(u=i|x)}>{\del S/\del p_{\txt{opt}}(u=j|x)}\right\}.
\ee
\end{prp}
\begin{IEEEproof}
We want to maximize $S$ with respect to $\vec{r}(x)=\big(p(u=0|x),...,p(u=M-1|x)\big)\in [0,1]^M$, with $\sum_{i=0}^{M-1} p(u=i|x) =1$ and $p(u=i|x)\geq 0$ for $i=0,\cdots, M-1$, i.e., $\vec{r}(x)$ is a point on the $M$-dimensional probability simplex
\be
\Delta_M(x)=\left\{\vec{r}(x)=\big(p(u=0|x),...,p(u=M-1|x)\big):\sum_{i=0}^{M-1}p(u=i|x)=1\right\}.\nn
%\Delta_M(x)=\left\{\vec{r}(x)=(p(u=0|x),...,p(u=M-1|x)):\sum_{i=0}^{M-1}p(u=i|x)=1\right\}.\nn
\ee
If $S$ is convex in $\vec{r}(x)$, then its maximum occurs at one or more corner points of $\Delta_M(x)$, i.e., $\vec{r}_{\txt{opt}}(x)=\vec{e}_j$ where  $\vec{e}_j=(0,...,0,\ub{1}_{j\txt{th spot}},0,...,0)$ for some $j\in \{0,...,M-1\}$. For each data point $x\in\X$, and any given point $\vec{r}(x)$ on the probability simplex $\Delta_M(x)$, let $\vec{t}_i(x)=\vec{e}_i-\vec{r}(x)$. Then by geometry of the graph of $S$ as a differentiable convex function of $\vec{r}(x)$, we observe that
  \be\ba
  \label{rg} &\vec{r}_{\txt{opt}}(x)=\vec{e}_i~\iff~p_{\txt{opt}}(u=i|x)=1\\
  &~~\iff~\txt{for all}~\vec{r}(x)\in \Delta_M(x)\backslash\{\vec{e}_i\},~~\vec{t}_i(x)\cdot{\del S/\del \vec{r}_{\txt{opt}}(x)}>0,
  \ea\ee
and,
  \be\ba
  &\label{rg2}\vec{r}_{\txt{opt}}(x)\neq\vec{e}_i~\iff~p_{\txt{opt}}(u=i|x)=0\\
  &~~\iff~\txt{for some}~\vec{r}(x)\in \Delta_M(x)\backslash\{\vec{e}_i\},~~\vec{t}_i(x)\cdot{\del S/\del \vec{r}_{\txt{opt}}(x)}<0,
  \ea\ee
where~ {\small$\del S/\del \vec{r}_{\txt{opt}}={\del S/\del \vec{r}}\big|_{\vec{r}=\vec{r}_{\txt{opt}}}$},~ $\vec{t}_i(x)\cdot{\del S/\del \vec{r}_{\txt{opt}}(x)}$ is the dot-product of the vectors $\vec{t}_i(x)$ and ${\del S/\del \vec{r}_{\txt{opt}}(x)}$, and $A\backslash B$ denotes the set difference, i.e., $A\backslash B=\{a: a\in A \mbox{ and } a \notin B\}$.

The region defined by $x$ satisfying $p_{\txt{opt}}(u=i|x)=1$ is
\be\ba
\label{dregion-proof}& R_{u=i}=\left\{x:~\vec{t}_i(x)\cdot{\del S/\del \vec{r}_{\txt{opt}}(x)}>0~~\txt{for all}~\vec{r}(x)\in\Delta_M(x)\backslash\{\vec{e}_i\}\right\}\\
&~~~~\sr{(a)}{=} \left\{x:~{\del S/\del p_{\txt{opt}}(u=i|x)}>{\del S/\del p_{\txt{opt}}(u=j|x)}~\txt{for all}~j\neq i \right\}\\
&~~~~=\bigcap_{j\neq i}\left\{x:~{\del S/\del p_{\txt{opt}}(u=i|x)}>{\del S/\del p_{\txt{opt}}(u=j|x)}\right\},
\ea\ee
which by (\ref{null_set}) is measure-wise complementary to the region defined by $x$ satisfying $p_{\txt{opt}}(u=i|x)=0$. Therefore, equation (\ref{rg}) covers both cases, and is equivalent to the deterministic rule (\ref{solution}). Note that step (a) in (\ref{dregion-proof}) is due to the following. For fixed $i$, let $\Delta_{M-1}^{(i)}(x)$ be the convex hull of $\{\vec{e}_j:~\txt{for all}~j\neq i\}$, which is the face of $\Delta_M(x)$ opposite to $\vec{e}_i$. Then the set of vectors
\be
\left\{\vec{t}_i(x)=\vec{e}_i-\vec{r}(x):~\vec{r}(x)\in \Delta_{M-1}^{(i)}(x)\right\}\nn
\ee
is the convex hull of $\{\vec{b}_j=\vec{e}_i-\vec{e}_j,~\txt{for all}~j\neq i\}$, i.e., for each $\vec{r}(x)\in \Delta_{M-1}^{(i)}(x)$, we can write $\vec{e}_i-\vec{r}(x)=\sum_{j\neq i}\al_j(\vec{e}_i-\vec{e}_j)$ for some nonnegative numbers $\al_j\geq0$ such that $\sum_j\al_j=1$. Finally, for any point $\vec{r}(x)\in\Delta_M(x)$, we can write
\bea
 \vec{r}(x)-\vec{e}_i=\big(\ld\vec{s}(x)+(1-\ld)\vec{e}_i\big)-\vec{e}_i=\ld\big(\vec{s}(x)-\vec{e}_i\big)\nn
\eea
for some $\ld\in[0,1]$ and some $\vec{s}(x)\in \Delta_{M-1}^{(i)}(x)$.
\end{IEEEproof}

The above proposition will be used in Sections \ref{section_NP} and \ref{section_KL} to determine optimal decision regions with the probability of detection and KL distance as objective functions. Before proceeding to the next section, we illustrate the key observations (\ref{rg}) and (\ref{rg2}) in the proof of the proposition with the simple case of binary decisions, followed by some remarks regarding applicability and possible extensions of the proposition.

\begin{figure}%[H] %\begin{figure}[htbp]
\centering
%\input{proposition1.pstex_t}% magnification = 37%
%\scalebox{2}{\input{proposition2.pstex_t}}% magnification = 37%
%\scalebox{2}{\input{proposition2.pdftex_t}}% magnification = 37%
%\scalebox{2}{\input{proposition3.pstex_t}}% magnification = 37%
\scalebox{2}{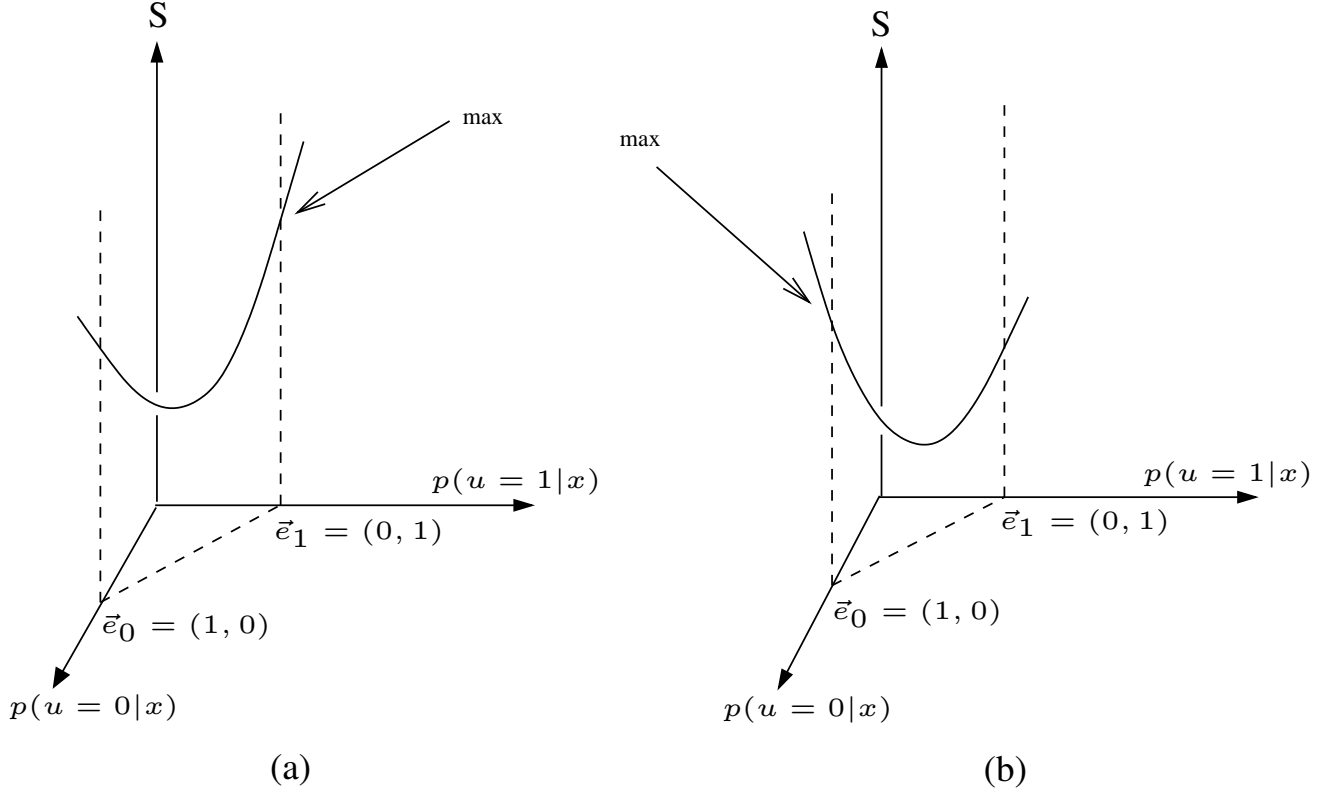}% magnification = 37%
\vspace{-0.2cm}
  \caption{Visualization of the objective function $S$ for $M=2$:~ In case (a), $x\in R_{u=1}$, and in case (b), $x\in R_{u=0}$.}\label{region-proof-diag}
\vspace{-0.4cm}
\end{figure}

\begin{comment}
We will verify in detail the equivalences (\ref{rg}) and (\ref{rg2}) for this special case, while noting that verification of the same equivalences for $M>2$ can be done by viewing the higher dimensional probability simplex $\Delta_M(x)$ as consisting of a collection of line segments along each of which the shape of the objective function $S$ is similar to its shape along $\Delta_2(x)$.
\end{comment}

Consider the binary case, i.e. $M=2$, in Fig. \ref{region-proof-diag}. We will verify in detail the equivalences (\ref{rg}) and (\ref{rg2}) for this special case, while noting that verification of the same equivalences for $M>2$ follows by noticing that the higher dimensional probability simplex $\Delta_M(x)$ can be viewed as a collection of line segments, and along the line segment from any point $\vec{r}(x)\in\Delta_M(x)$ to any one of the corner points $\{\vec{e}_i\}$ of $\Delta_M(x)$, the shape (of the graph) of the objective function $S$ is similar to its shape along $\Delta_2(x)$. This similarity in shape is due to the fact that a function is convex if and only if it is convex along each line segment through its domain. For the convex function $S$ depicted in Fig. \ref{region-proof-diag}, it is apparent that in case $(a)$ the optimum occurs at $\vec{r}_{\rm{opt}}=\vec{e}_1=(0,1)$, and in case (b) the optimum occurs at $\vec{r}_{\rm{opt}}=\vec{e}_0=(1,0)$. Let $\vec{t}_i(x)=\vec{e}_i-\vec{r}(x),~i=1,0$, and
\be
\Delta_2(x)=\left\{\big(p(u=0|x),p(u=1|x)\big):~p(u=0|x)+p(u=1|x)=1\right\}.\nn
\ee
For a given $x\in \X$, suppose that we have the situation in Fig. \ref{region-proof-diag}(a), $i.e.,\vec{r}_{\txt{opt}}(x)=\vec{e}_1$. It is straightforward to check that for all $\vec{r}(x)\in \Delta_2(x)\backslash\{\vec{e}_1\}$, the projection $(\vec{e}_1-\vec{r}(x))\cdot\del S/\del \vec{r}(x)\big|_{\vec{r}(x)=\vec{e}_1}$ of the slope $\del S/\del \vec{r}(x)\big|_{\vec{r}(x)=\vec{e}_1}$ along the direction $\vec{e}_1-\vec{r}(x)$ is positive. However, since only the direction of the vector $\vec{e}_1-\vec{r}(x)$ is relevant, it suffices to consider only the point $\vec{r}(x)=\vec{e}_0=(1,0)$, in which case we have that
\be\ba
&(\vec{e}_1-\vec{e}_0)\cdot{\del S\over\del \vec{r}(x)}\big|_{\vec{r}(x)=\vec{e}_1}={\del S\over\del p(u=1|x)}\big|_{p(u=1|x)=1}-{\del S\over\del p(u=0|x)}\big|_{p(u=0|x)=0}\nn\\
&~={\del S\over\del p(u=1|x)}\big|_{p(u=1|x)=p_{\txt{opt}}(u=1|x)}-{\del S\over\del p(u=0|x)}\big|_{p(u=0|x)=p_{\txt{opt}}(u=0|x)}\nn\\
&~>0,~~\Ra~~x\in R_{u=1}.\nn
\ea\ee
This explains the forward implication $\Ra$ in equivalence (\ref{rg}). The reverse implication $\La$ in the same equivalence is also true for the following reason. Suppose
\bea
\label{optimal-derivative}\big(\vec{e}_1-\vec{r}(x)\big)\cdot{\del S\over\del\vec{r}_{\txt{opt}}(x)}>0~~\txt{for each}~~\vec{r}(x)\in\Delta_2(x)\backslash\{\vec{e}_1\}.
\eea
By convexity of $S$, for any $\al\in (0,1)$,
\be
S\big(\al \vec{r}(x)+(1-\al)\vec{r}_{\txt{opt}}(x)\big)\leq \al S\big(\vec{r}(x)\big)+(1-\al)S\big(\vec{r}_{\txt{opt}}(x)\big),\nn
\ee
which in the limit $\al\ra 0$ implies
\be
\label{optimal-derivative-convexity}\big(\vec{r}_{\txt{opt}}(x)-\vec{r}(x)\big)\cdot {\del S\over\del \vec{r}_{\txt{opt}}(x)}\geq S\big(\vec{r}_{\txt{opt}}(x)\big)-S\big(\vec{r}(x)\big).
\ee
With $\vec{r}_{\txt{opt}}(x)=\vec{e}_1$, the above inequality (\ref{optimal-derivative-convexity}) is consistent with (\ref{optimal-derivative}), i.e., the LHS of (\ref{optimal-derivative-convexity}) has a consistently nonnegative sign. However, the same is not true with $\vec{r}_{\txt{opt}}(x)=\vec{e}_0$, which completes the verification of the reverse implication $\La$ in (\ref{rg}).
\begin{comment}
By the intermediate value theorem of calculus,

{\footnotesize
\bea
0\leq S(\vec{r}_{\txt{opt}}(x))-S(\vec{r}(x))=\big(\vec{r}_{\txt{opt}}(x)-\vec{r}(x)\big)\cdot {\del S\over\del\vec{r}_{\txt{int}}(x)}
\eea
}for some point $\vec{r}_{\txt{int}}(x)\in [\vec{r}_{\txt{opt}}(x),\vec{r}(x)]$.
 \end{comment}
Noting that statement (\ref{rg2}) is simply the negation of statement (\ref{rg}) up to the zero probability data sets (\ref{null_set}), the complementary picture in Fig. \ref{region-proof-diag}(b) explains the complementary equivalence (\ref{rg2}) in the same way.

\begin{rmks}~
\begin{enumerate}%[leftmargin=0.5cm]%[leftmargin=*]
\item~ The binary decision rule can be simplified further. In this case, the probability simplex $\Delta_2(x)$ is the single line with equation $p(u=0|x)+p(u=1|x)=1$. Thus by the chain rule of differentiation, the differential operator $\vec{t}_i(x)\cdot\del /\del \vec{r}_{\txt{opt}}(x)$ along $\Delta_2(x)$ is equivalent to a derivative, which we denote by $\del^B /\del p_{\txt{opt}}(u=i|x)$, with the property
\bea
\label{binary-derivative}{\del^B p(u|x)/\del p(u'|x')}=(-1)^{u-u'}\delta_{xx'}
\eea
in addition to linearity and the Leibnitz rule. Hence the binary decision regions take the compact form
\bea
\label{binary-decision-regions}R_{u=i}=\left\{x:~{\del^B S/\del p_{\txt{opt}}(u=i|x)}>0 \right\},
\eea
where the superscript $B$ in $\del^B$ serves as a reminder to the reader of the property (\ref{binary-derivative}) which ensures that the derivative is restricted to the probability simplex $\Delta_2(x)$. The compact form of the binary decision rule as implemented by (\ref{binary-derivative}) and (\ref{binary-decision-regions}) will greatly simplify calculations later on.

\item~ Notice that (\ref{solution}) is an implicit equation in $p_{\txt{opt}}(u=i|x)$ since the region $R_{u=i}$ also depends on $p_{\txt{opt}}(u=i|x)$. Therefore we must proceed to substitute the equations \be
    \left\{p_{\txt{opt}}(u=i|x)=I_{R_{u=i}}(x): i=0,1,...,M-1\right\}\nn
    \ee
    into the objective function $S$, and then compute the optimal threshold values that explicitly determine the decision regions. In the case of distributed networks of sensors where more than one set of local decision rules are involved, the resulting system of equations is often analytically intractable and one has to resort to numerical computation.

\item~ Recall that the set of data points satisfying (\ref{null_set}) must be null with respect to the probability measure. Otherwise, the deterministic rule (\ref{solution}) is replaced by a randomized version
\bea
\label{solution-randomized}p_{\txt{opt}}(u=i|x)=I_{R_{u=i}}(x)+\sum_k\rho_{ik}I_{C_k}(x),
\eea
where $\{C_k\}$ is a partition of the set $\bigcup_iC_{u=i}$ and $\rho_{ik}\in [0,1],~\sum_i\rho_{ik}=1,$ are arbitrary (i.e., free) coefficients but which must be consistent with all the constraints of the optimization problem. It is worthwhile to remark that the deterministic rule (\ref{solution}) is more easily realized when $x$ is continuous than when $x$ is discrete. Thus, the randomized rule (\ref{solution-randomized}) is often required when $x$ is a discrete random variable.

\item~ Since the Bayes risk is an affine (hence convex/concave) function of $p(u|x)$, Proposition \ref{region_proposition} is a direct generalization of the familiar procedure whereby the unconditional Bayes risk
\bea
R(\gamma)=\sum_{i,j}C_{ij}p(\gamma(x)=i,H_j)=\sum_{i,x}p(\gamma(x)=i|x)R_i(x),
\eea
is minimized over $\gamma$ simply by separately minimizing the associated conditional Bayes risks $R_i(x)=\sum_jC_{ij}p(x|H_j)p(H_j)$ over $i$ by means of the choice
 \bea
 p(\gamma(x)=i|x)=\left\{
                    \begin{array}{ll}
                      1, & R_i(x)<R_j(x)~\txt{for all}~j\neq i, \\
                      0, & \txt{otherwise},
                    \end{array}
                  \right.\nn
 \eea
where we have assumed for simplicity that $P\big(R_i(x)=R_j(x)\big)=0$ for all $i$ and $j\neq i$.
Note that our method of optimization is different from the optimal control theoretical approach considered in \cite{TPK}. Using standard methods of convex optimization, Proposition \ref{region_proposition} can be further extended to include non-differentiable convex objective functions by replacing the derivative $\del S/\del p(u=i|x)$ with a subdifferential.
% The proposition can equally be adapted for some classes of harmonic functions that include convex functions as special cases.
\end{enumerate}
\end{rmks}

For ease of presentation, we consider the case of $M=2$ in Sections \ref{section_NP} and \ref{section_KL}, i.e., binary hypotheses with binary  sensor output. For the same reason, we will consider only cases where the optimal decision rule is deterministic as stated in Proposition \ref{region_proposition}. Generalizations to that of multi-level sensor outputs as well as to systems involving multiple sensors will be described in Section V.

%\newpage
\section{The fixed sample size Neyman-Pearson test}\label{section_NP}
We will now consider an NP test with observations that can be of arbitrary but fixed dimension, i.e., the test is a fixed sample size NP test. From Fig. \ref{one-and-two-way-diagram}(a) the decisions $(v,w)$ for the YX process are based on the observations $(x,y)$, which satisfy the conditional independence relation~ $p_i(x,y)=p_i(x)p_i(y).$~ Similarly Fig. \ref{one-and-two-way-diagram}(b) describes the XYX process with decisions $(u,v,w)$.
\begin{comment}
We will now consider an NP test using $x$. Notice $x$ itself can be of arbitrary but fixed dimension -- the test is a fixed sample size NP test. From Fig. \ref{one-and-two-way-diagram}(a) the decisions $(v,w)$ for the YX process are based on the observations $(x,y)$, which satisfy the conditional independence relation~ $p_i(x,y)=p_i(x)p_i(y).$~ Similarly Fig. \ref{one-and-two-way-diagram}(b) describes the XYX process with decisions $(u,v,w)$.
\end{comment}

Since we have more than one decision in a decentralized sensor network, we also have more than one optimization variable. Thus, to apply Proposition \ref{region_proposition} to the YX process for example, we must first replace $p(u|x)$ by the variables $p(v|y)$ and $p(w|x,v)$. We then optimize the objective function by separately applying Proposition \ref{region_proposition} to each variable (whilst the other variables are held constant) to obtain the system of deterministic decision rules
\bea
\label{decentralized-rule1}&&p_{\txt{opt}}(v=1|y)=I_{R_{v=1}}(y),\\
\label{decentralized-rule2}&&p_{\txt{opt}}(w=1|x,v)=I_{R_{w=1|v}}(x),
\eea
which is a coupled system of equations at the overall optimal point $\big(p_{\txt{opt}}(v|y),p_{\txt{opt}}(w|x,v)\big)$. By Proposition II.1, if the deterministic decision rules (\ref{decentralized-rule1}) and (\ref{decentralized-rule2}) are based on a convex objective function $S$, then
\bea
&&R_{v=1}=\left\{y:{\del^B S}/\del p_{\txt{opt}}(v=1|y)>0\right\},\nn\\
&&R_{w=1|v}=\left\{x:{\del^B S}/\del p_{\txt{opt}}(w=1|x,v)>0\right\}.\nn
\eea
We will use the above optimization technique for all problems in this section and in subsequent sections. At this point, we would like to remark that using this optimization technique, the decision/fusion rules that underly the work of Zhu et al in \cite{zhu-et.al1,zhu-et.al2} readily follow from Proposition \ref{region_proposition}.

The objective for the NP test is the maximization of the probability of detection whilst the probability of false alarm must not exceed a certain fixed value $\al$:
\bea
\label{NP-problem}&& \txt{maximize}~~~~ P_d = p_1(w=1)\nn\\
&&\txt{subject to}~~~~ P_f =
p_0(w=1)\leq \al
\eea
The Lagrangian for this optimization problem is given by
\bea
\label{NP-Lagrangian-0} L=p_1(w=1)+\ld~\big(\al-p_0(w=1)\big), ~~\ld\geq 0.
\eea
For the YX process, $L$ is a function of $\ld$, $p(v|y)$, and $p(w|x,v)$, and
\bea
\label{NP-YX-probabilty}p_i(w)=\sum_{x,y,v}p(w|x,v)p(v|y)~p_i(x)p_i(y),
\eea
while for the XYX process, $L$ is a function of $\ld$, $p(u|x)$, $p(v|y,u)$, and $p(w|x,v)$, and
\bea
\label{NP-XYX-probabilty}p_i(w)=\sum_{x,v,y,u}p(w|x,v)p(v|y,u)p(u|x)~p_i(x)p_i(y).
\eea
Furthermore, for the YX process, we need to minimize $L$ over  $\ld$ in the dual problem, or we can equivalently solve $p_0(w=1)=\al$ for $\ld(\al)$ since $p_1(w=1)$ and $p_0(w=1)$ are linear and have the same monotonicity properties as functions of $p(w=1|x,v)$ and $p(v=1|y)$. Similarly for the XYX process, we need to minimize $L$ over  $\ld$ in the dual problem, or we can equivalently solve $p_0(w=1)=\al$ for $\ld(\al)$ since $p_1(w=1)$ and $p_0(w=1)$ are linear and have the same monotonicity properties as functions of $p(w=1|x,v)$,~ $p(v=1|y,u)$, and $p(u=1|x)$.

The fact that the XYX process performs at least as well as the YX process is trivial: any detection performance achieved by a YX process can be achieved by an XYX process that simply ignores the first decision variable $u$. It remains to show that there are cases where the involvement of the initial decision variable $u$, i.e., the interactive fusion, strictly improves upon the one-way tandem fusion.

Notice that the Lagrangian (\ref{NP-Lagrangian-0}) has essentially the same form as a Bayes risk function such as the probability of error used in \cite{s-a-c}, the only difference being the presence of the Lagrange multiplier $\ld$ as a new optimization variable.

%%%%%%%%%%%%%%%%%%%%%%%%%%%%%%%%%%%%%%%%%%%%%%%%%%%%%%%%%%%%%%%%%%%%%%
\begin{thm}\label{np-theorem}
The NP test with objective given by (\ref{NP-Lagrangian-0}) has the following decision regions. For the YX process, we have
\be\ba
\label{NP-YX-regions}R_{v=1}=\left\{y:~{p_1(y)\over p_0(y)}>\ld^{(2)}\right\},~~~~R_{w=1|v}=\left\{x:~{p_1(x)\over p_0(x)}>\ld_{v}^{(3)}\right\},
\ea\ee
where $\ld^{(2)}=\ld~{P_0(R_{w=1|v=1})-P_0(R_{w=1|v=0})\over P_1(R_{w=1|v=1})-P_1(R_{w=1|v=0})}$ and $\ld_{v}^{(3)}=\ld~{P_0(R_{v})\over P_1(R_{v})}$.
For the XYX process, we have
\be\ba
\label{NP-XYX-regions}& R_{u=1}=\left\{x:~{p_1(x)\over p_0(x)}Q(x)>\ld^{(1)}Q(x)\right\},\\
&R_{v=1|u}=\left\{y:~{p_1(y)\over p_0(y)}>\ld^{(2)}_u\right\},\\
&R_{w=1|v}=\left\{x:~{p_1(x)\over p_0(x)}>\sum_u\ld_{vu}^{(3)}~I_{R_u}(x)\right\},
\ea\ee
where  $\ld^{(1)}=\ld~{P_0(R_{v=1|u=1})-P_0(R_{v=1|u=0})\over P_1(R_{v=1|u=1})-P_1(R_{v=1|u=0})}$, $Q(x)=I_{R_{w=1|v=1}}(x)-I_{R_{w=1|v=0}}(x)$,\\ $\ld^{(2)}_u=\ld~{P_0(R_{w=1|v=1}\cap R_u)-P_0(R_{w=1|v=0}\cap R_u)\over P_1(R_{w=1|v=1}\cap R_u)-P_1(R_{w=1|v=0}\cap R_u)}$, and $\ld_{vu}^{(3)}=\ld~{P_0(R_{v|u})\over P_1(R_{v|u})}$.
\end{thm}
\begin{IEEEproof}
 See APPENDIX \ref{NP-test-details}.
\end{IEEEproof}
\begin{rmks}~
\begin{enumerate}%[leftmargin=0.5cm]%[leftmargin=*]
\item[5)]~ Note that the compact form of the decision rules in Theorem \ref{np-theorem} includes the case in which the denominator of $\ld^{(1)}$, $\ld^{(2)}$, or $\ld^{(2)}_u$ approaches zero, i.e., infinite threshold values are not meaningless. In other words, the decision rules as stated in Theorem \ref{np-theorem} incur no inconsistency or loss of generality as long as the thresholds are allowed to take infinitely large values (e.g., when a threshold denominator approaches zero). There are indeed cases where the objective function (as a function of the thresholds) attains its optimum only when one or more of these thresholds approach infinity. This corresponds to the degenerate decision regions where the decision is independent of the observation (i.e., the output of the decision rule is a constant). Alternatively, one can circumvent the issue of diminishing denominator by rewriting the decision rule in a less compact form; The way to do this explicitly is apparent from the intermediate steps in the proof of the theorem in APPENDIX \ref{NP-test-details}. The above remarks will apply to the asymptotic test as well.
\item[6)]~ While it is apparent that for the YX process, both decision rules amount to a LRT, the same is not true for the XYX process. This is because of the dependence introduced in the interactive fusion scheme: while the observations $x$ and $y$ are themselves conditionally independent, the two step interaction introduces conditional dependence between $x$ and $v$ due to the initial input $u$ from X to Y. The situation here is similar to that found in \cite{s-a-c} as well as for the ``Unlucky Broker Problem'' of \cite{MMM1,MMM2}, where the observations are conditionally independent but the decision rules are not determined by simple likelihood ratio tests.
\end{enumerate}
\end{rmks}
The Lagrangian in (\ref{NP-Lagrangian-0}) can be equivalently expressed in terms of the obtained decision regions in Theorem \ref{np-theorem}, i.e.,
\be\ba
L=\sum_{v}P_1(R_{v})P_1(R_{w=1|v}) +\ld~\left[\al-\sum_{v}P_0(R_{v})P_0(R_{w=1|v})\right]
\ea\ee
for the YX Process, and
\be\ba
L=\sum_{u,v}P_1(R_{v|u})P_1(R_{w=1|v}\cap R_u)+\ld\left[\al-\sum_{u,v}P_0(R_{v|u})P_0(R_{w=1|v}\cap R_u)\right]\nn
\ea\ee
for the XYX Process.

%%%%%%%%%%%%%%%%%%%%%%%%%%%%%%%%%%%%%%%%%%%%%%%%%%%%%%%%%%%%%%
%\subsection{A Gaussian Example}
\subsection{Example: Constant Signal in White Gaussian Noise}
\begin{figure}%[H]
\centering
\includegraphics[width=15cm,height=9cm]{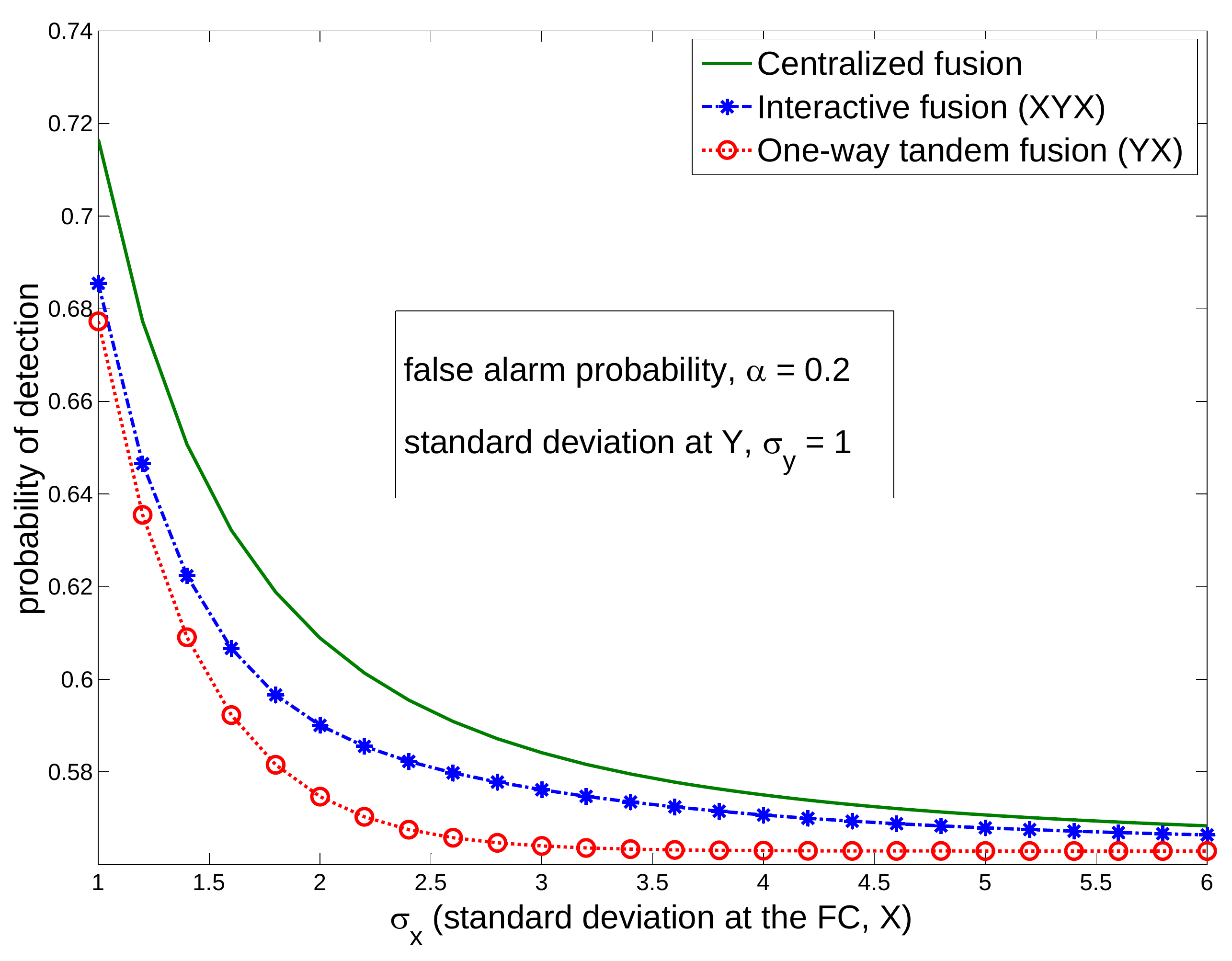}\\
\vspace{-0.2cm}
  \caption{Performance of XYX and YX processes: These graphs were generated via numerical iteration following the analysis carried out in Section \ref{section_NP} and Appendix \ref{wgn-details}.}\label{np-graph}
\vspace{-0.4cm}
\end{figure}

We now use a simple example to show that the XYX process can strictly outperform the YX process for fixed sample NP test. Consider the detection of a constant signal $s$ in white Gaussian noise with observations
\bea
\label{WGN}x=s+z_1,~~~~y=s+z_2,
\eea
where $z_1\sim N(0,\sigma_x^2),~~z_2\sim N(0,\sigma_y^2)$ and $z_1$ and $z_2$ are independent of each other, and the two hypotheses under test are
\bea
H_0:~s=0,~~~~H_1:~s=1.\nn
\eea
After simplifying assumptions for the XYX process (see APPENDIX \ref{wgn-details}), the decision regions for this example take the following simple forms. For YX,
\be\ba
& R_{v=1}=\left\{y>t^{(2)}=\sigma^2_y\log\ld^{(2)}+1/2\right\},\nn\\
& R_{w=1|v}=\left\{x>t^{(3)}_v=\sigma_x^2\log\ld^{(3)}_v+1/2\right\},\nn
\ea\ee
and for XYX,
\be\ba
& R_{u=1}=\left\{x>t^{(1)}=\sigma^2_x\log\ld^{(1)}+1/2\right\},\nn\\
& R_{v=1|u}=\left\{y>t^{(2)}_u=\sigma_y^2\log\ld^{(2)}_u+1/2\right\},\nn\\
& R_{w=1|v}=\bigcup_uR_{w=1|v,u},\nn\\ &R_{w=1|v,u}=\left\{x>t^{(3)}_{vu}=\sigma^2_x\log\ld^{(3)}_{vu}+1/2\right\},\nn
\ea\ee
where the thresholds (i.e., the $\ld$'s) are as in (\ref{NP-YX-regions}) and (\ref{NP-XYX-regions}), and we have assumed for simplicity that each sensor's observation consists of only one real sample, i.e., $x,y\in\Real=\X=\Y$. Fig. \ref{np-graph} shows the dependence of the probability of detection on $\sigma_x$ when $\sigma_y$ is fixed. The corresponding false alarm probability  $P_f=0.2$. Thus, the XYX process has strictly larger probability of detection compared with the YX process.

The curve corresponding to centralized fusion in Fig. \ref{np-graph} is obtained by repeating the same optimization procedure using (\ref{NP-problem}) and (\ref{NP-Lagrangian-0}), but with the probability of the centralized decision $w=\rho(x,y)$ given by $p_i(w=1)=\sum_{x,y}p(w=1|x,y)p_i(x,y)$. Here, the decision rule $p(w=1|x,y)=I_{R_{w=1}}(x,y)$, the constant false alarm probability constraint $\al=p_0(w=1)$, and the detection probability $P_d=p_1(w=1)$ can be easily written as
\be\ba
& R_{w=1}=\left\{(x,y): {x\over\sigma_x^2}+{y\over\sigma_y^2}>t=\ln\ld+{1\over 2\sigma_x^2}+{1\over 2\sigma_y^2}\right\},\\
& \al=\int_{-\infty}^\infty Q\left(\sigma_yt-{\sigma_y\over\sigma_x}{x\over\sigma_x}\right){e^{-{x^2\over 2\sigma_x^2}}\over\sqrt{2\pi\sigma_x^2}}dx,\\
& P_d=\int_{-\infty}^\infty Q\left(\sigma_yt-{\sigma_y\over\sigma_x}~{x\over\sigma_x}-{1\over\sigma_y}\right){e^{-{(x-1)^2\over 2\sigma_x^2}}\over\sqrt{2\pi\sigma_x^2}}dx,\nn
\ea\ee
where the threshold $t$ as a function of $\al$ is obtained by solving the constant false alarm probability constraint.

%\newpage
\section{The asymptotic Neyman-Pearson test}\label{section_KL}
Let us consider $n$ observation samples~ $(x_1,y_1),...,(x_n,y_n)$, and suppose processing is carried out on a sample-by-sample basis. Using the XYX process as an illustration, the two sensors go through, for each $k=1,\cdots,n$, a decision process with $u_k=\gamma_k(x_k),~~v_k=\delta_k(y_k,u_k)$. The final decision at node X utilizes the entire observation sequence $x^n$ and the output sequence $v^n$ from node Y, i.e., $w=\rho(x^n,v^n)$. We have, therefore,
\bea
p_i(w)=\sum_{x^n,v^n}p(w|x^n,v^n)p_i(x^n,v^n),\nn
\eea
where $p(w|x^n,v^n)$ is determined by the final decision rule.

Using Proposition \ref{region_proposition} we obtain the decision region for $w$ as
\bea
 R_{w=1}=\mathop{\txt{arg max}}_{\substack{R\subset \X^n\times\{0,1\}^n,\\p_0(R)\leq\al}}p_1(R)=\left\{(x^n,v^n):{p_1(x^n,v^n)\over p_0(x^n,v^n)}>\ld\right\},
\eea
where $\ld$ is the Lagrange multiplier as in (\ref{NP-Lagrangian-0}). We could proceed by Proposition \ref{region_proposition} to find decision regions for $v^n$.

However, we are interested mainly in the asymptotic detection performance, i.e., when $n$ grows large. We emphasize here that the processing at node Y is memoryless, i.e., $v_k$ is only a function of the current observation $y_k$. As such, given that $y_k$ is an i.i.d. sequence and that the decision rule $\gamma(\cdot)$ for each $k$ is identical, the pairs $(x_k,v_k)$ form an i.i.d. sequence. Recall that by the weak law of large numbers, under $H_0$,
\be\ba
& {1\over n}\log{p_0(x^n,v^n)\over p_1(x^n,v^n)}={1\over n}\sum_{k=1}^n\log{p_0(x_k,v_k)\over p_1(x_k,v_k)}\nn\\
&~~\sr{n\ra\infty}{\ral}~ E_{p_0(x,v)}\log{p_0(x,v)\over p_1(x,v)}=D(p_0(x,v)\|p_1(x,v)),\nn
\ea\ee
where $v=\delta(y,u),~~u=\gamma(x).$~ By the Chernoff-Stein Lemma, \cite{cover-thomas}, the test with acceptance region for $H_0$
\be
R^n_\vep(p_0|p_1)=\left\{(x^n,v^n):D(p_0\|p_1)-\vep\leq {1\over n}\log{p_0(x^n,v^n)\over p_1(x^n,v^n)}\leq D(p_0\|p_1)+\vep\right\},\nn
\ee
is asymptotically optimal, with error exponent
\bea
\label{kl-distance}-\lim_{n\ra\infty}{1\over n}\log p_1\big((x^n,v^n)\in R^n_\vep(p_0|p_1)\big)= D(p_0(x,v)\|p_1(x,v)),
\eea
which is the KL distance
%\footnote{Compare with Chernoff information $C[x,v]=-\min_{0\leq \ld\leq 1}\log ~\sum_{x,v}~p_1(x,v)~e^{\ld\log{p_0(x,v)\over p_1(x,v)}}$}
that we will now use as our objective function for the asymptotic performance of the NP test.

We show in the following that with the KL distance as objective, interactive fusion provides no improvement over one-way tandem fusion. Therefore, for large sample size~ $n\ra\infty$,~ interactive fusion does not perform better than one-way tandem fusion.

\subsection{One-way tandem fusion (YX process)}\label{one-pass}
In the one-way tandem fusion network, as illustrated in Fig. \ref{one-and-two-way-diagram}(a), Y makes a decision $v=\delta(y)$ and passes $v$ to X. The optimal decision $v$ is chosen so as to maximize the KL distance
\bea
&&K[x,v]=D\big(p_0(x,v)\|p_1(x,v)\big)
\eea
 at sensor X.

Since ~$p_i(x,v)=p_i(x)p_i(v),$ we have
\be
\label{KL-YX} K[x,v]=D\big(p_0(x)\|p_1(x)\big)+\sum_vp_0(v)~\log\big(p_0(v)/ p_1(v)\big)
\ee
where {~$p_i(v)=\sum_yp(v|y)p_i(y).$}
\begin{thm}\label{XY_theorem}
The optimal decision region at Y is given by
\bea
\label{omit1}&&R_{v=1}=\bigg\{y:{p_1(y)\over p_0(y)}>\ld\bigg\},\\
\label{YX-threshold}&&\ld=\bigg(\log{\beta(1-\al)\over\al(1-\beta)}\bigg)\bigg/\bigg({\beta-\al\over \beta(1-\beta)}\bigg),
\eea
where $\al = P_0\left({p_1(y)/p_0(y)}>\ld\right)$ and $\beta = P_1\left({p_1(y)/p_0(y)}>\ld\right).$ Thus $\al$, $\beta$, and $\ld$ are coupled with each other.
\end{thm}
\begin{IEEEproof}
The key observation is that the KL distance (\ref{KL-YX}) is differentiable and convex in $p(v|y)$. This follows because the KL distance is jointly convex in $\big(p_0(v),p_1(v)\big)$ while $p_0(v)$ and $p_1(v)$ are each affine functions of $p(v|y)$. Thus Proposition \ref{region_proposition} applies.
In addition, because of the constraint { $p(v=1|y)+p(v=0|y)=1$}, the derivative of { $K[x,v]\triangleq K^{\txt{YX}}$} requires the differentiation rule
\bea
\label{diff-rule}&&{\del^B p(v|y)/\del p(v'|y')}=(-1)^{v-v'}\delta_{yy'},
\eea
which holds for binary decisions. It remains only to evaluate the derivative of $K[x,v]$, which we do in the following.
\begin{equation}
     \begin{aligned}
&{\del^B K[x,v]\over \del p(v=1|y)}={\del^B \over \del p(v=1|y)}\bigg[\sum_v~\sum_{y'}p(v|{y'})p_0({y'})~\log{\sum_{y'}p(v|{y'})p_0({y'})\over \sum_{y'}p(v|{y'})p_1({y'})}\bigg]\nn\\
&~~\sr{(a)}{=}\sum_v~\sum_{y'}(-1)^{v-1}\delta_{yy'}p_0({y'})~\log{p_0(v)\over p_1(v)}+\sum_vp_0(v)\left[{\sum_{y'}(-1)^{v-1}\delta_{yy'}p_0({y'})\over p_0(v)}-{\sum_{y'}(-1)^{v-1}\delta_{yy'}p_1({y'})\over p_1(v)}\right]\nn\\
%&&~~\sr{(aa)}{=}\sum_v(-1)^{v-1}p_0({y})~\log{p_0(v)\xover p_1(v)}+\sum_vp_0(v)\left[{(-1)^{v-1}p_0({y})\over p_0(v)}-{(-1)^{v-1}p_1({y})\over p_1(v)}\right]\\
&~~\sr{(b)}{=}p_0({y})\sum_v(-1)^{v-1}\log{p_0(v)\over p_1(v)}-p_1({y})\sum_v(-1)^{v-1}{p_0(v)\over p_1(v)}\nn\\
&~~\sr{(c)}{=}-p_0({y})\log{p_1(v=1)[1-p_0(v=1)]\over p_0(v=1)[1-p_1(v=1)]}+p_1({y}){p_1(v=1)-p_0(v=1)\over p_1(v=1)[1-p_1(v=1)]}\nn\\
&~~\sr{(d)}{=}-p_0(y)\log{\beta(1-\al)\over\al(1-\beta)}+p_1(y){\beta-\al\over \beta(1-\beta)}.\nn
     \end{aligned}
\end{equation}
Step $(a)$ follows by the chain rule of differentiation and (\ref{diff-rule}).  The last but one term at step $(a)$ is zero, leading to the simplification at step $(b)$. It is already apparent at this stage that we have a likelihood ratio test. Step $(c)$ is obtained after summing over $v$ and carrying out some elementary rearrangements. Finally, at step $(d)$ we make the substitution $\al=p_0(v=1)$ which is equal to $P_0(R_{v=1}){\triangleq} P_0\left({p_1(y)/p_0(y)}>\ld\right)$, and similarly $\beta=p_1(v=1),$ which is equal to $P_1(R_{v=1}){\triangleq} P_1\left({p_1(y)/p_0(y)}>\ld\right)$. On application of Proposition \ref{region_proposition}, the condition ${\del K[x,v]/\del p_{\txt{opt}}(v=1)>0}$ leads directly to (\ref{omit1}) and (\ref{YX-threshold}).
\end{IEEEproof}

Notice that the decision region defined in (\ref{omit1}) and the threshold of the likelihood ratio of $y$ given in (\ref{YX-threshold}) are coupled with each other. Iterative process is thus needed for finding the optimal $\lambda$ and the associated $\alpha$ and $\beta$. An alternative is to directly obtain the optimal values for the thresholds as those that optimize the objective as a function of the thresholds, which typically involves exhaustive search but is immune to the issue of local optimum.

The maximum KL distance is given by
\be
\label{KYX0}K^{\txt{YX}}_{\max}=K[x]+\al^\ast\log{\al^\ast \over\beta^\ast}+(1-\al^\ast)\log{1-\al^\ast\over 1-\beta^\ast},
\ee
where $K[x]=D(p_0(x)\|p_1(x))$ and $\al^\ast$ and $\beta^\ast$ are the values of $\al$ and $\beta$ that maximize the KL distance.

We now revisit the hypothesis test described in (\ref{WGN}).
By Theorem \ref{XY_theorem},~ the optimal decision region at $Y$ is  $R_{v=1}=\left\{y:~y>t=\sigma^2_y\log\ld(t)+{1\over 2}\right\}$,~ where
\bea
\ld(t)=\left(\log{Q\left({t-1\over\sigma_y}\right)\left(1-Q\left({t\over\sigma_y}\right)\right)\over Q\left({t\over\sigma_y}\right)\left(1-Q\left({t-1\over\sigma_y}\right)\right)}\right)\bigg/\left({Q\left({t-1\over\sigma_y}\right)-Q\left({t\over \sigma_y}\right)\over Q\left({t-1\over\sigma_y}\right)\left(1-Q\left({t-1\over\sigma_y}\right)\right)}\right).\nn
\eea
The corresponding maximum KL distance is
\be
K_{\max}[x,v]={1\over 2\sigma_x^2}+Q\left({t^{\ast}\over\sigma_y}\right) \log{Q\left({t^{\ast}\over\sigma_y}\right) \over Q\left({t^{\ast}-1\over\sigma_y}\right) }+\left(1-Q\left({t^{\ast}\over\sigma_y}\right)\right)\log{1-Q\left({t^{\ast}\over\sigma_y}\right) \over 1-Q\left({t^{\ast}-1\over\sigma_y}\right) },\nn
\ee
where $t^\ast$ is the threshold that maximizes the KL distance. While the decision is still in the (equivalent) form of an LRT, this threshold is different from that of the fixed sample size test.

 %That is, while both fixed-sample test and KL case have similar expressions, the actual thresholds used are different since the lagrangian multipliers are different.

\subsection{Interactive fusion (XYX process)}\label{two-pass}
For interactive fusion illustrated in Fig. \ref{one-and-two-way-diagram}(b), X makes a decision $u=\gamma(x)$ and passes $u$ onto Y. Y further makes a decision $v=\delta(y,u)$ and sends it back to X. The optimal decisions $u$ and $v$ are chosen so as to maximize the KL distance $K[x,v]{\triangleq} K^{\txt{XYX}}$ in the final step at X. The KL distance can be written as
\be\ba
\label{KL-XYX}K^{\txt{XYX}}=D\big(p_0(x,v)\|p_1(x,v)\big)=D\big(p_0(x)\|p_1(x)\big)+\sum_xp_0(x)\sum_vp_0(v|x)~\log{p_0(v|x)\over p_1(v|x)},
\ea\ee
where $p_i(v|x)=\sum_{u}p(u|x)\sum_yp(v|y,u)p_i(y)$. Different from the YX process, $v$ is now conditionally dependent of X as $v$ takes $u=\gamma(x)$ as its input.

\begin{thm}\label{XYX_theorem}
 For the XYX process, the optimal decision region at sensor X is given by
\bea
\label{omit2}&R_{u=1}=\left\{x:\sum_uI_{R_u}(x)A_uB_u>0\right\},\\
\label{X-thresholds}&~~~~~~~~ A_u={\beta^{(2)}_u-\al^{(2)}_u\over \beta^{(2)}_u(1-\beta^{(2)}_u)},~~~~B_u={\beta^{(2)}_1-\beta^{(2)}_0\over \al^{(2)}_1-\al^{(2)}_0}-\ld^{(2)}_u,
\eea
and the optimal decision regions at sensor Y are given by
\bea
\label{omit3}& R_{v=1|u}=\bigg\{y:{p_1(y)\over p_0(y)}>\ld^{(2)}_u\bigg\},\hspace{0.5cm}&\\
\label{Y-thresholds}&~~\ld^{(2)}_u=\bigg(\log{\beta^{(2)}_u(1-\al^{(2)}_u)\over\al^{(2)}_u(1-\beta^{(2)}_u)}\bigg)\bigg/\bigg({\beta^{(2)}_u-\al^{(2)}_u\over \beta^{(2)}_u(1-\beta^{(2)}_u)}\bigg),
\eea
where { $\al^{(2)}_u=P_0(R_{v=1|u})$}, and {$\beta^{(2)}_u=P_1(R_{v=1|u})$}.
\end{thm}
The proof requires the following lemma whose proof follows immediately from the properties of a partition of a given set.
\begin{lmm}\label{partition-lemma}
Let $\{R_i:~i=1,...,m\}$ be any partition of the data space $\X$. Then for any continuous multivariate function $f$, and for each $x\in\X$,
\bea
\label{identity1}f\left(\sum_{i=1}^mI_{R_i}(x)a_{i1},\sum_{i=1}^mI_{R_i}(x)a_{i2},...\right)=\sum_{i=1}^mI_{R_i}(x)~f\big(a_{i1},a_{i2},...\big),
\eea
where $a_{ij}$ for all $i$ and $j$ are numbers.
\begin{comment}
When the set $\{R_i:~i=1,...,m\}$ is not a partition, the simple function $g_j=\sum_iI_{R_i}(x)a_{ij}$ can always be rewritten in a separable form
\be\ba
&g_j=\sum_iI_{R_i}(x)a_{ij}=\sum_l\sum_iI_{R_i\cap\td{R}_l}(x)a_{ij}\nn\\
&~~~~\eqv \sum_lI_{\td{R}_l}(x)\td{a}_{lj},
\ea\ee
where $\{\td{R}_l:~l=1,...,m'\}$ is a partition of $\X$.
\end{comment}
\end{lmm}

\begin{IEEEproof}[Proof of Theorem \ref{XYX_theorem}]
The KL distance (\ref{KL-XYX}) is differentiable and convex in $p(v|y,u)$ and $p(u|x)$, and thus Proposition \ref{region_proposition} applies. With this observation, we only need to evaluate the derivatives of the KL distance with respect to $p(v=1|y,u)$ and $p(u=1|x)$.

For the region $R_{u=1}$ at sensor $X$, we have

{\small
%\fontsize{4.5}{6}\selectfont
\be\ba
&{\del^B K[x,v]\over\del p(u=1|x) }={\del^B\over\del p(u=1|x) }\left[\sum_{x'}p_0({x'})\sum_vp_0(v|{x'})~\log{p_0(v|{x'})\over p_1(v|{x'})}\right]\\
&~~\sr{(a)}{=}\sum_{x',v}p_0({x'})\left({\del^B p_0(v|{x'})\over\del p(u=1|x) }~\log{p_0(v|{x'})\over p_1(v|{x'})}+p_0(v|{x'})~\left[{{\del^B p_0(v|{x'})\over\del p(u=1|x)}\over p_0(v|{x'})}-{{\del^B p_1(v|{x'})\over\del p(u=1|x)}\over p_1(v|{x'})}\right]\right)\\
&~~\sr{(b)}{=}\sum_{x',v,u}p_0({x'})(-1)^{u-1}\delta_{xx'}\left(P_0(R_{v|u})~\log{p_0(v|{x'})\over p_1(v|{x'})}+p_0(v|{x'})~\left[{P_0(R_{v|u})\over p_0(v|{x'})}-{P_1(R_{v|u})\over p_1(v|{x'})}\right]\right)\\
%&~~\sr{}{=}p_0({x})\sum_{v,u}(-1)^{u-1}\left(P_0(R_{v|u})~\log{p_0(v|{x})\over p_1(v|{x})}+p_0(v|{x})~\left[{P_0(R_{v|u})\over p_0(v|{x})}-{P_1(R_{v|u})\over p_1(v|{x})}\right]\right)\\
&~~\sr{(c)}{=}p_0({x})\sum_{v,u}(-1)^{u-1}\left(P_0(R_{v|u})~\log{p_0(v|{x})\over p_1(v|{x})}-P_1(R_{v|u}){p_0(v|{x})\over p_1(v|{x})}\right)\\
&~~\sr{(d)}{=}p_0({x})\sum_{v,u'}(-1)^{u'-1}\left(P_0(R_{v|u'})~\log{\sum_uI_{R_u}(x)P_0(R_{v|u})\over \sum_uI_{R_u}(x)P_1(R_{v|u})}-P_1(R_{v|u'}){\sum_uI_{R_u}(x)P_0(R_{v|u})\over \sum_uI_{R_u}(x)P_1(R_{v|u})}\right)\\
&~~\sr{(e)}{=}p_0({x})\sum_{v,u}I_{R_u}(x)\left(\sum_{u'}(-1)^{u'-1}P_0(R_{v|u'})~\log{P_0(R_{v|u})\over P_1(R_{v|u})}-\sum_{u'}(-1)^{u'-1}P_1(R_{v|u'}){P_0(R_{v|u})\over P_1(R_{v|u})}\right).\nn
\ea\ee
}$(a)$ is from the chain rule for differentiation, $(b)$ is due to the differentiation rule (\ref{diff-rule}), the last but one term at step $(b)$ is zero - leading to $(c)$, $(d)$ is expansion by chain rule for probabilities showing dependence on the indicator functions of the decision regions (due to (\ref{optimal1})) and finally, $(e)$ follows from $(d)$ by the identity (\ref{identity1}).

Thus we obtain the region given by (\ref{omit2}). Similarly, for the regions $R_{v=1|u}$ in (\ref{omit3}), the same steps as above apply as follows.

{\small
 %\fontsize{4.5}{6}\selectfont
\be\ba
&{\del^B K[x,v]\over\del p(v=1|y,u) }={\del^B\over\del p(v=1|y,u) }\left[\sum_xp_0(x)\sum_vp_0(v|x)~\log{p_0(v|x)\over p_1(v|x)}\right]\\
&~~\sr{(a)}{=}\sum_xp_0(x)\sum_v\left({\del^B p_0(v|x)\over\del p(v=1|y,u) }~\log{p_0(v|x)\over p_1(v|x)}+p_0(v|x)~\left[{{\del^B p_0(v|x)\over\del p(v=1|y,u)}\over p_0(v|x)}-{{\del^B p_1(v|x)\over\del p(v=1|y,u)}\over p_1(v|x)}\right]\right)\\
&~~\sr{(b)}{=}\sum_xp_0(x)p(u|x)\sum_v(-1)^{v-1}\left(p_0(y)~\log{p_0(v|x)\over p_1(v|x)}+p_0(v|x)~\left[{p_0(y)\over p_0(v|x)}-{p_1(y)\over p_1(v|x)}\right]\right)\\
&~~\sr{(c)}{=}\sum_xp_0(x)p(u|x)\sum_v(-1)^{v-1}\left(p_0(y)~\log{p_0(v|x)\over p_1(v|x)}-p_1(y){p_0(v|x)\over p_1(v|x)}\right)\\
&~~\sr{(d)}{=}\sum_xp_0(x)I_{R_u}(x)\sum_v(-1)^{v-1}\left(p_0(y)~\log{\sum_{u'}I_{R_{u'}}(x)P_0(R_{v|{u'}})\over \sum_{u'}I_{R_{u'}}(x)P_1(R_{v|{u'}})}-p_1(y){\sum_{u'}I_{R_{u'}}(x)P_0(R_{v|{u'}})\over \sum_{u'}I_{R_{u'}}(x)P_1(R_{v|{u'}})}\right)\nn\\
&~~\sr{}{=}\sum_{x}p_0(x)I_{R_u}(x)\sum_{u'}I_{R_{u'}}(x)\sum_v(-1)^{v-1}\left(p_0(y)~\log{P_0(R_{v|{u'}})\over P_1(R_{v|{u'}})}-p_1(y){P_0(R_{v|{u'}})\over P_1(R_{v|{u'}})}\right)\\
&~\sr{(e)}{=}\!\sum_{x}p_0(x)I_{R_u}(x)\sum_v(-1)^{v-1}\!\left(p_0(y)~\log{P_0(R_{v|u})\over P_1(R_{v|u})}-p_1(y){P_0(R_{v|u})\over P_1(R_{v|u})}\right)\\
&~~=P_0(R_u)\left[p_0(y)\sum_v(-1)^{v-1}\log{P_0(R_{v|u})\over P_1(R_{v|u})}-p_1(y)\sum_v(-1)^{v-1}{P_0(R_{v|u})\over P_1(R_{v|u})}\right]\\
&~~=P_0(R_u)\left[-p_0(y)\log{\beta^{(2)}_u(1-\al^{(2)}_u)\over\al^{(2)}_u(1-\beta^{(2)}_u)}+p_1(y){\beta^{(2)}_u-\al^{(2)}_u\over \beta^{(2)}_u(1-\beta^{(2)}_u)}\right].
\ea\ee
}Thus we obtain likelihood ratio tests at $Y$.
\end{IEEEproof}

Using (\ref{identity1}), the KL distance (\ref{KL-XYX}) can be expressed as
\bea
&& K^{\txt{XYX}}=K[x]+\sum_{u,v}P_0(R_u)~P_0(R_{v|u})\log{P_0(R_{v|u})\over P_1(R_{v|u})}\nn\\
\label{KXYX0}&&~~~~=K[x]+\al^{(1)}f(\al_1^{(2)},\beta_1^{(2)})+(1-\al^{(1)})f(\al_0^{(2)},\beta_0^{(2)}),\nn\\
\eea
where~ $ K[x]=D(p_0(x)\|p_1(x)),$~  $\al^{(1)}$ is a constant independent of the thresholds, and
\bea
\label{KL-function}f(\al,\beta)=\al \log{\al \over \beta }+(1-\al )\log{1-\al \over 1-\beta }.
\eea
Thus we have the following theorem.

\begin{prp}\label{XYX-trivial-region}~
 The YX and XYX processes achieve identical $K[x,v]$. That is,
 \bea
 K^{\txt{YX}}_{\max}= K^{\txt{XYX}}_{\max}.
 \eea
\end{prp}
\begin{IEEEproof}
The KL distances achieved by the two fusion systems, $K^{\txt{YX}}$ from (\ref{KYX0}) and $K^{\txt{XYX}}$ from (\ref{KXYX0}), are respectively
\bea
\label{KYX}K^{\txt{YX}}&=&K[x]+f(\alpha,\beta)\\
\label{KXYX}K^{\txt{XYX}}&=&K[x]+\al^{(1)}f(\al_1^{(2)},\beta_1^{(2)})+(1-\al^{(1)})f(\al_0^{(2)},\beta_0^{(2)}),
\eea
where the function $f(\al,\beta)$ is defined in (\ref{KL-function}).

Let $\alpha^*$ and $\beta^*$ be the optimal values that maximize $f(\al,\beta)$ in $K^{\txt{YX}}$. Comparing  (\ref{omit1})-(\ref{YX-threshold}) and  (\ref{omit3})-(\ref{Y-thresholds}), it is apparent that the same $\alpha^*$ and $\beta^*$ also maximize both $f(\alpha_1^{(2)},\beta_1^{(2)})$ and $f(\alpha_0^{(2)},\beta_0^{(2)})$ in $K^{\txt{XYX}}$. This is so since for each value of $u$, the threshold dependence of the LRT using $y$ is identical to that used in the YX process. Thus, the optimal decision on $v$ at Y for the XYX process simply ignores the input from $u$, leading to identical LRTs for both values of $u$.

\begin{comment}
Comparing $K^{YX}$ from (\ref{KYX0}) and $K^{XYX}$ from (\ref{KXYX0}), we have
\bea
\label{KYX}K^{YX}&=&K[x]+f(\alpha,\beta)\\
\label{KXYX}K^{XYX}&=&K[x]+\al^{(1)}f(\al_1^{(2)},\beta_1^{(2)})+(1-\al^{(1)})f(\al_0^{(2)},\beta_0^{(2)}),
\eea
where the function $f$ is given by (\ref{KL-function}). Let $\alpha^*$ and $\beta^*$ be the optimal values that maximize the threshold dependent term $f(\alpha,\beta)$ in $K^{YX}$. From  (\ref{KYX}) and  (\ref{KXYX}), it is apparent that the same $\alpha^*$ and $\beta^*$ also maximize the threshold dependent terms $f_0=\al^{(1)}f(\al_0^{(2)},\beta_0^{(2)})$ and $f_1=(1-\al^{(1)})f(\al_1^{(2)},\beta_1^{(2)})$ in $K^{XYX}$, since $\al^{(1)}$ is constant (i.e., independent of thresholds) and these two terms depend on separate threshold variables. Thus, the optimal decision $v$ for Y simply ignores the input $u$ from X. Hence XYX achieves the same performance as YX.
\end{comment}
\end{IEEEproof}

\begin{figure}%[H]
\centering
\includegraphics[width=15cm,height=9cm]{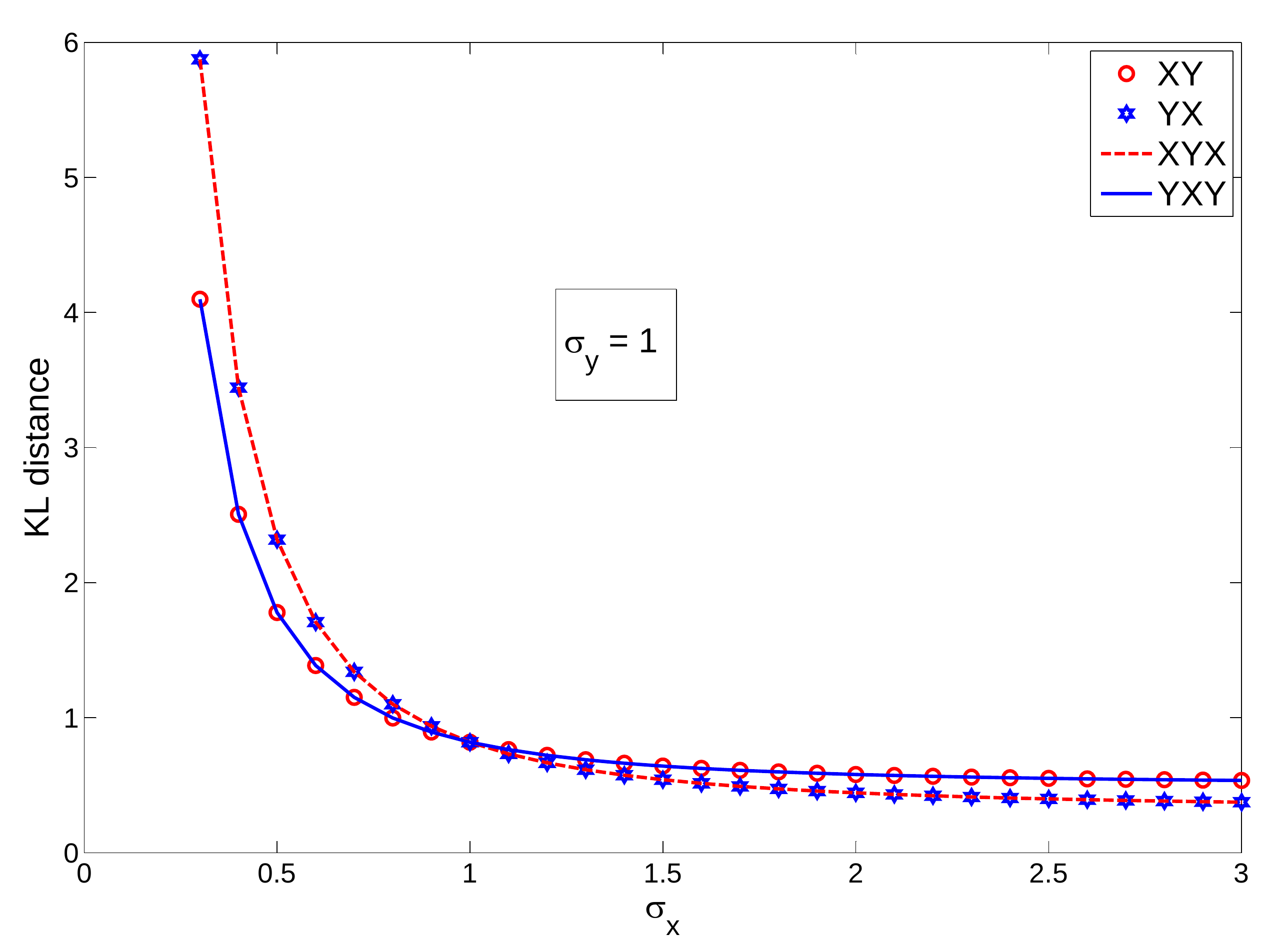}\\
\vspace{-0.2cm}
  \caption{Comparison of KL distances of one-way tandem fusion and interactive fusion with different communication directions.  For this plot, we fix $\sigma_y=1$ throughout while varying $\sigma_x$.}\label{kl-graph}
\vspace{-0.2cm}
\end{figure}

Proposition \ref{XYX-trivial-region} holds for any probability distribution. The results for the constant signal in WGN under hypotheses (\ref{WGN}) are shown in Fig. \ref{kl-graph}, where the KL distances of YX and XYX processes coincide with each other. Also plotted are the KL distances of XY and YXY that also coincide with each other. An interesting observation from the plot is that the two sets of curves, each corresponding to making final decision at different nodes, intercept each other at the point when $\sigma_x=\sigma_y=1$. Thus for this example, it is always better to make the final decision at the sensor with better signal to noise ratio.

%\begin{comment}

\section{Generalizations} \label{generalize}
We have shown that while interactive fusion may strictly improve the detection performance of fixed sample size NP test, there is no advantage for the large sample test as the one-way fusion achieves exactly the same error exponent. The result was derived under a two-sensor system model with a single round of interaction and $1$-bit sensor output. We now generalize the result to more realistic settings that may involve multiple round of iterations involving multiple sensors and soft (i.e., multi-bit) sensor output.
\subsection{Multiple-step memoryless interactive fusion (MIF)}\label{N-step process}
\begin{figure}%[H]
\centering
%\scalebox{2}{\input{fusion-diagrams-multiple-step.pstex_t}}% magnification = 50%
\scalebox{2}{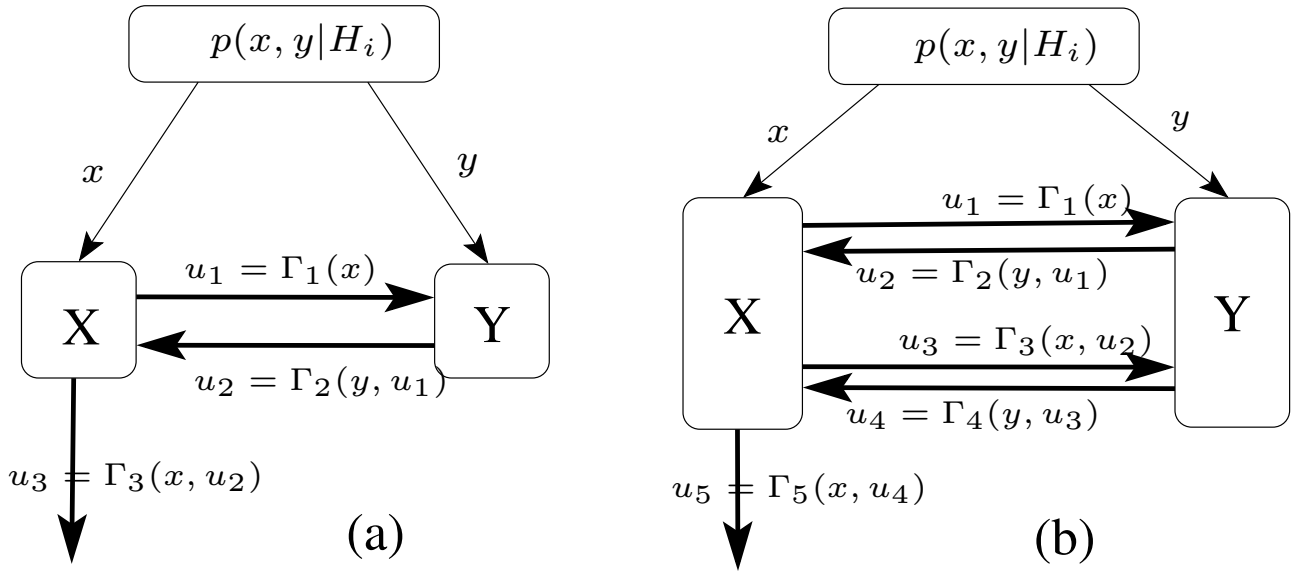}% magnification = 50%
\vspace{-0.2cm}
\caption{Sample MIF processes: (a) $N=3$~ MIF (XYX process), and (b) $N=5$~ MIF (XYXYX process).}\label{fusion-diagrams-multiple-step}
\vspace{-0.4cm}
\end{figure}

In multiple round interactive fusion, sensors exchange $1$-bit information iteratively in $N>3$ steps. It is not difficult to show that this multiple round interactive fusion may strictly improve that of the one-way tandem fusion in its asymptotic performance, i.e., it achieves a higher KL distance compared with that of the one-way tandem fusion. We note that an $N$-round interactive fusion should perform at least as well as that of one-way fusion where the output from sensor $Y$ has $N$ bits. It is apparent that a one-way tandem fusion with $N$ bits may strictly outperform that with a single bit. Indeed, for $N$ large enough, the performance will approach that of the centralized detection.

However, there might be situations where the multiple round interactive fusion may proceed in a memoryless fashion, which we refer to as \emph{memoryless interactive fusion} (MIF). That is, the sensor output depends only on the previous input from the other sensor as well as its own observation. We show that for this memoryless processing model, multiple-step interactive fusion has no advantage in its asymptotic detection performance over the one-way tandem fusion.

We begin with the expansion of the probability~ $p_i(u_N)=p(u_N|H_i)$~ of the final decision $u_N$. Denote any sequence $s_1,...,s_N$ by $s^N$. Let $u^N$ be the sequence of decisions in the MIF process XYXY$\cdots$YX involving two independent sensors X and Y, and let
\bea
\label{N-step-pdata}z^N\eqv(z_1,...,z_N)=(x,y,x,y,...,y,x)
\eea
be the corresponding sequence of observations used at processing, as shown in Fig. \ref{fusion-diagrams-multiple-step} for $N=3,5$. Here we assume $N$ is odd, thus the decision process always starts with and ends at node $X$. Then using $u_k=\Gamma_k(z_k,u_{k-1}),$ and that $z_k=x$ when $k$ is odd, and $z_k=y$ when $k$ is even, we obtain
\be\ba
\label{odd-case1} &p_i(u_N)=\sum_{z^N,u^{N-1}}p_i(z^N)\prod_{k=1}^Np(u_k|z_k,u_{k-1})\\
&~~=\sum_{x,y,u^{N-1}}p_i(x,y)\prod_{r=1}^{(N-1)/2}\left[p(u_{2r-1}|x,u_{2r-2})p(u_{2r}|y,u_{2r-1})\right].
\ea\ee
Now using $p_i(x,y)=p_i(x)p_i(y),$ we have
\be
\label{odd-case2}   p_i(u_N|x)=\sum_{y,u^{N-1}}p_i(y)\prod_{r=1}^{(N-1)/2}\left[p(u_{2r-1}|x,u_{2r-2})p(u_{2r}|y,u_{2r-1})\right].
\ee
Based on this expansion of $p_i(u_N)$, the following lemma (proved in Appendix \ref{proof-N-step-lemma}) gives the peculiar nature of the resulting decision regions that are determined by an observation that is directly involved in the KL distance.

\begin{lmm}[Degenerate MIF decision regions]\label{N-step-lemma}
Let $u_N=\Gamma_N(x,u_{N-1})$ be the decision at the final step of a MIF process XYXY$\cdots$YX with independent observations $x$ and $y$. Let the objective function be given by the KL distance at the final step
\bea
\label{N-step-KL-1} K[x,u_{N-1}]=\sum_{x,u_{N-1}}p_0(x,u_{N-1})\log{p_0(x,u_{N-1})\over p_1(x,u_{N-1})}.
\eea
Then all decision regions based on $x$, with decisions ~$u_{2r-1}=\Gamma_{2r-1}(x,u_{2r-2}),~r=1,2,...,{N-1\over 2}$, have the following general form.
\bea
\label{trivial_regions} R_{u_{2r-1}=1|u_{2r-2}}=\left\{x:~\sum_{\al}I_{D_{\al}}(x) A_{\al,r,u_{2r-2}}>0\right\},
\eea
where $\{D_\al\}$ is a partition of the data space $\X$, and the coefficients $A_{\al,r,u_{2r-2}}$ are independent of $x$.
\end{lmm}

Notice that (\ref{omit2}) is a special case of (\ref{trivial_regions}). The following are some remarks about the degenerate decision regions (\ref{trivial_regions}):
\bit%[leftmargin=0.5cm]%[leftmargin=*]
\item They depend on the distributions $p_0(x)$ and $p_1(x)$ only globally over $\X$, and not pointwise in $x$. Therefore given a single data point $x\in\X$, they cannot distinguish between $H_0$ and $H_1$.
\item They are determined by piecewise constant functions with discrete probability distributions, and hence cannot define independent continuous threshold parameters; i.e., they contain no independent thresholds.
\item They have piecewise constant probability; i.e., have same probability under both hypotheses.
\item Their only role is to reparametrize the thresholds of the other regions. Consequently, they cannot improve optimality of the KL distance (as the next lemma shows).
\eit
The following lemma shows that the decision regions given by (\ref{trivial_regions}) are trivial in the sense that they do not participate in the decision process.
\begin{lmm}\label{covergence_lemma}
With respect to dependence on thresholds, the decision regions (\ref{trivial_regions}) of Lemma \ref{N-step-lemma} have piece-wise constant probability measures. Moreover, such probability measures play no role at convergence and therefore do not contribute to the overall decision process.
\end{lmm}

\begin{IEEEproof}
Under $H_i$, the probability $P_i\left(R_{u_{2r-1}=1|u_{2r-2}}\right)$ of the region $R_{u_{2r-1}=1|u_{2r-2}}$ is given by
\be\ba
\label{region-probability1}& P_i\left(\sum_{\al}~I_{D_{\al}}(x) A_{\al,r,u_{2r-2}}>0\right)=\sum_x p_i(x)~I\left(\sum_{\al}I_{D_{\al}}(x) A_{\al,r,u_{2r-2}}>0\right)\\
&~~~~\sr{(\ref{identity1})}{=}\sum_x p_i(x)\sum_{\al}~I_{D_{\al}}(x)~ I\left(A_{\al,r,u_{2r-2}}>0\right)=\sum_{\al}~P_i(D_{\al})~I\left(A_{\al,r,u_{2r-2}}>0\right).
\ea\ee
On the other hand, we can expand the probabilities of the decision regions $\{R_{u_{2r-1}=1|u_{2r-2}}:r=1,...,(N-1)/2\}$ as follows. For each $r$ we have some subset $\{\al_r\}\subset\{\al\}$ of the index set $\{\al\}$ such that
\be\ba
\label{region-probability2}&P_i\left(R_{u_{2r-1}=1|u_{2r-2}}\right)=P_i\left(I_{R_{u_{2r-1}=1|u_{2r-2}}}(x)=1\right)=P_i\left(\sum_{\al_r\in\{\al_r\}}I_{D_{\al_r}}(x)=1\right)\\
&~~~~=\sum_{\al_r}P_i(D_{\al_r}).
\ea\ee
Hence equating (\ref{region-probability1}) and (\ref{region-probability2}) for each $r\in \{1,...,(N-1)/2\}$, we can solve the resulting linear system of equations for the values of $\{P_i(D_\al)\},$ from which it is clear that $P_0(D_\al)=P_1(D_\al)$ for each $\al$, and that the probabilities are piecewise constant in the thresholds even if the system of equations is underdetermined (i.e., has more unknowns than equations). Therefore, the probabilities $P_i\left(R_{u_{2r-1}=1|u_{2r-2}}\right)$ are also piecewise constant with respect to threshold dependence at X.

Ultimately to have convergence in the iteration, $P_i\left(R_{u_{2r-1}=1|u_{2r-2}}\right)$ must settle on one of its possible constant values. That is, $P_i\left(R_{u_{2r-1}=1|u_{2r-2}}\right)$ is ultimately constant in the iteration process. Thus beyond what initial conditions can achieve,~ $\{P_i\left(R_{u_{2r-1}=1|u_{2r-2}}\right)\}$~ do not play any role in determining the point at which convergence occurs.
\end{IEEEproof}

Therefore, careful analysis of the MIF process shows that whenever a sensor's data is explicitly summed over in the KL distance, the decision process becomes independent of that particular sensor's data. Since repetition of the decision process involving only one sensor's data cannot improve performance, it follows that MIF processing does not improve performance with respect to the KL distance.

\subsection{Interactive fusion between the FC and K sensors}\label{multiple-sensors}
\begin{figure}%[H]
\centering
%\input{fusion-diagrams-multiple-sensor.pstex_t}% magnification = 50%
%\caption{Sample MIF processes with $K$ peripheral sensors: (a) $N=3$ multi-sensor MIF (X$\vec{\txt{Y}}$X process), and (b) $N=5$ milti-sensor MIF (X$\vec{\txt{Y}}$X$\vec{\txt{Y}}$X process).}\label{fusion-diagrams-multiple-sensor}
%\scalebox{2}{\input{fusion-diagrams3-multiple-sensor.pstex_t}}% magnification = 50%
\scalebox{2}{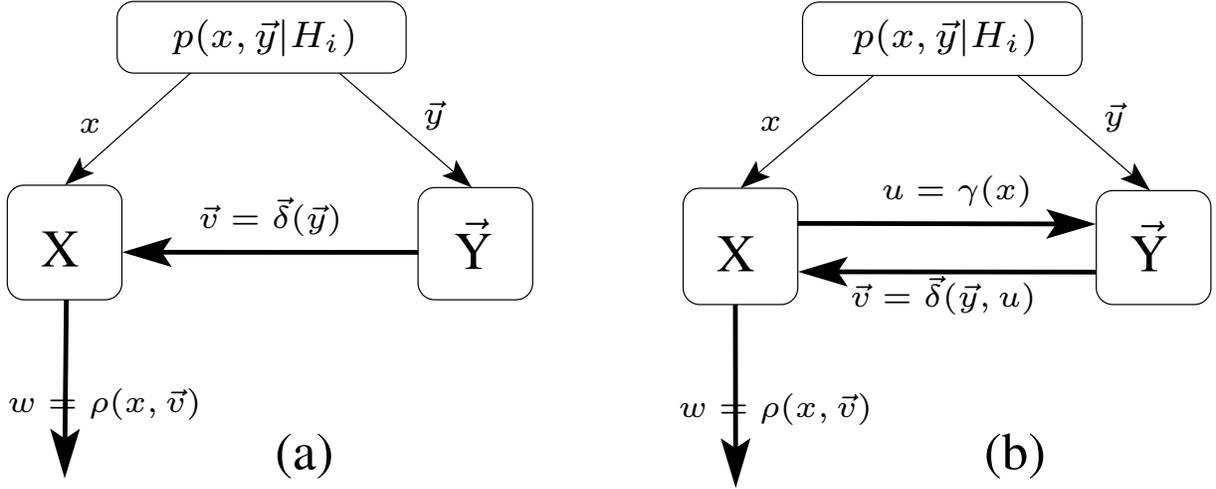}% magnification = 50%
\vspace{-0.2cm}
\caption{(a) = one-way tandem fusion ($\vec{\txt{Y}}$X process), and (b) = interactive fusion (X$\vec{\txt{Y}}$X process)}\label{fusion-diagrams3-multiple-sensor}
\vspace{-0.4cm}
\end{figure}

Consider our main setup in Fig. \ref{one-and-two-way-diagram} and maintain sensor X as the FC while replacing sensor $Y$ by K different sensors $\vec{\txt{Y}}=\{\txt{Y}_1,...,\txt{Y}_K\}$, with respective independent observations $\vec{y}=\{y_1,...,y_K\}$. The resulting system is shown in Fig. \ref{fusion-diagrams3-multiple-sensor}. For the $\vec{\txt{Y}}$X process, we have decisions $(\vec{v},w)\eqv(v_1,...,v_K,w)$ based on observations $(x,\vec{y})\eqv(x,y_1,...,y_K)$, where $\vec{v}=\vec{\delta}(\vec{y})=\big(\delta_1(y_1),...,\delta_K(y_K)\big)$ and $w=\rho(x,\vec{v})\eqv \rho(x,v_1,...,v_K)$. Similarly, in the X$\vec{\txt{Y}}$X process, the decisions $(u,\vec{v},w)\eqv(u,v_1,...,v_K,w)$ are based on observations $(x,\vec{y})\eqv(x,y_1,...,y_K)$, with $u=\gamma(x)$, $\vec{v}=\vec{\delta}(\vec{y},u)=\big(\delta_1(y_1,u),...,\delta_K(y_K,u)\big)$ and $w=\rho(x,\vec{v})\eqv \rho(x,v_1,...,v_K)$.

In the fixed sample size NP test with Lagrangian (\ref{NP-Lagrangian-0}), $p_i(w)$ is given by
\be\ba
p_i(w)=\sum_{x,{\vec{y}},{\vec{v}}}p(w|x,{\vec{v}})p({\vec{v}}|{\vec{y}})~p_i(x)p_i({\vec{y}})=\sum_{x,{\vec{y}},{\vec{v}}}p(w|x,{\vec{v}})~\prod_{k=1}^Kp(v_k|y_k)~p_i(x)~\prod_{k=1}^Kp_i(y_k)
\ea\ee
for the $\vec{\txt{Y}}$X process, and
\be\ba
&p_i(w)=\sum_{x,{\vec{v}},{\vec{y}},u}p(w|x,{\vec{v}})p({\vec{v}}|{\vec{y}},u)p(u|x)~p_i(x)p_i({\vec{y}})\\
&~~~~=\sum_{x,{\vec{v}},{\vec{y}},u}p(w|x,{\vec{v}})~\prod_{k=1}^Kp(v_k|y_k,u)~p(u|x)~p_i(x)~\prod_{k=1}^Kp_i(y_k)
\ea\ee
for the X$\vec{\txt{Y}}$X process.
It suffices to find the X$\vec{\txt{Y}}$X decision regions only since those for $\vec{\txt{Y}}$X can be deduced from them by simply deleting the first decision $u$. Using Proposition \ref{region_proposition}, and with the same steps as in the proof of Theorem \ref{np-theorem} in APPENDIX \ref{NP-test-details}, we obtain the following.
\be\ba
R_{u=1}=\left\{x:~{\del^B S\over\del p(u=1|x)}>0\right\}=\left\{x:~{p_1(x)\over p_0(x)}Q(x)>\ld^{(1)}Q(x)\right\},
\ea\ee
where $\ld^{(1)}=\ld^{(0)}~{\prod_{k=1}^KP_0(R_{v_k=1|u=1})-\prod_{k=1}^KP_0(R_{v_k=1|u=0})\over \prod_{k=1}^KP_1(R_{v_k=1|u=1})-\prod_{k=1}^KP_1(R_{v_k=1|u=0})},$~ $\ld^{(0)}$ is still given by (\ref{ld0}),
 \\ {$Q(x)=-\sum_{\vec{v}}(-1)^{v_1+...+v_K}I_{R_{w=1|\vec{v}}}(x)$}, and the objective function $S$ is given by (\ref{uniform-objective}).

For each $k=1,...,K$,
\be\ba
 R_{v_k=1|u}=\left\{y_k:~{\del^B S\over\del p(v_k=1|y_k,u)}>0\right\}=\left\{y_k:~{p_1(y_k)\over p_0(y_k)}>\ld^{(2)}_{k,u}\right\},
\ea\ee
where
\be
\ld^{(2)}_{k,u}=\ld^{(0)}~{\sum_{{v\backslash v_k}}\left[P_0\left(R_{w=1|{v\backslash v_k},v_k=1}\cap R_u\right)-P_0\left(R_{w=1|{v\backslash v_k},v_k=0}\cap R_u\right)\right]\prod_{k'\neq k}P_0\left(R_{v_{k'}|u}\right) \over \sum_{{v\backslash v_k}}\left[P_1\left(R_{w=1|{v\backslash v_k},v_k=1}\cap R_u\right)-P_1\left(R_{w=1|{v\backslash v_k},v_k=0}\cap R_u\right)\right]\prod_{k'\neq k}P_1\left(R_{v_{k'}|u}\right) }.\nn
\ee
Similarly, at $X$,
\be\ba
\label{XYX-region-33} &R_{w=1|{\vec{v}}}=\left\{x:~{\del^B S\over\del p(w=1|x,{\vec{v}})}>0\right\}=\left\{x:~{p_1(x)\over p_0(x)}>\sum_u\ld_{{\vec{v}}u}^{(3)}~I_{R_u}(x)\right\},
\ea\ee
where { $\ld_{{\vec{v}}u}^{(3)}=\ld^{(0)}~{\prod_{k=1}^KP_0\left(R_{v_k|u}\right)\over \prod_{k=1}^K P_1\left(R_{v_k|u}\right)}$}.

\begin{figure}%[H]
\centering
%\scalebox{2}{\input{fusion-diagrams-multiple-sensor.pstex_t}}% magnification = 50%
\scalebox{2}{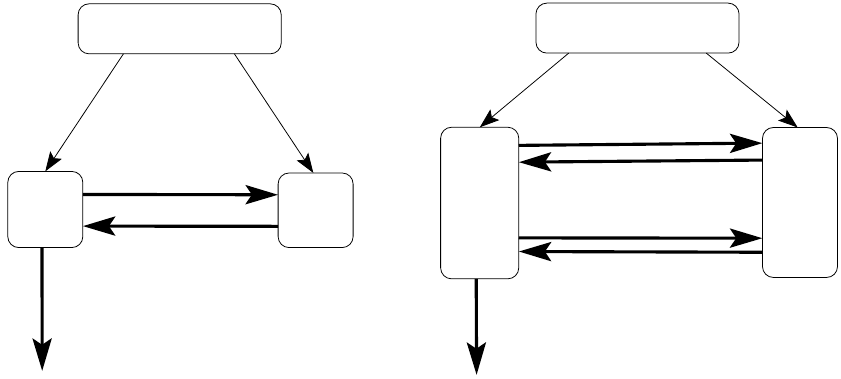}% magnification = 50%
\vspace{-0.2cm}
\caption{Sample MIF processes with $K$ peripheral sensors: (a) $N=3$ multi-sensor MIF (X$\vec{\txt{Y}}$X process), and (b) $N=5$ milti-sensor MIF (X$\vec{\txt{Y}}$X$\vec{\txt{Y}}$X process).}\label{fusion-diagrams-multiple-sensor}
\label{fusion-diagrams-multiple-sensor2}
\vspace{-0.4cm}
\end{figure}

Similarly, for the asymptotic Neyman-Pearson test, the KL distance {\small $K^{\txt{X$\vec{\txt{Y}}$X}}\triangleq K[x,\vec{v}]=D\big(p_0(x,{\vec{v}})\|p_1(x,{\vec{v}})\big)$} can be expressed as
\bea
D\big(p_0(x,{\vec{v}})\|p_1(x,{\vec{v}})\big)=D\big(p_0(x)\|p_1(x)\big)+\sum_xp_0(x)\sum_{\vec{v}}p_0({\vec{v}}|x)~\log{p_0({\vec{v}}|x)\over p_1({\vec{v}}|x)},
\eea
where $p_i({\vec{v}}|x)=\sum_{u}p(u|x)\sum_yp({\vec{v}}|y,u)p_i(y)$. Through the same steps as in the proof of Theorem \ref{XYX_theorem} for the X$\vec{\txt{Y}}$X process, the decision region at sensor X is
\be\ba
& R_{u=1}=\left\{x:{\del^B K[x,\vec{v}]\over\del p(u=1|x) }>0\right\}=\left\{x:\sum_uI_{R_u}(x)C_{u}>0\right\},\\
&C_{u}=\sum_{\vec{v}}\bigg(\sum_{u'}(-1)^{u'-1}P_0(R_{{\vec{v}}|u'})~\log{P_0(R_{{\vec{v}}|u})\over P_1(R_{{\vec{v}}|u})}-\sum_{u'}(-1)^{u'-1}P_1(R_{{\vec{v}}|u'}){P_0(R_{{\vec{v}}|u})\over P_1(R_{{\vec{v}}|u})}\bigg),
\ea\ee
and the pair of decision regions $R_{v=1|u}$ at sensor $Y$ has the following $K$ analogues corresponding to the sensors $\vec{\txt{Y}}$; for each $k=1,...,K,$
\be\ba
& R_{v_k=1|u}=\bigg\{y_k:~{\del^B K[x,{\vec{v}}]\over\del p(v_k=1|y_k,u) }>0\bigg\}=\bigg\{y_k:{p_1(y_k)\over p_0(y_k)}>\ld^{(2)}_{ku}\bigg\},\hspace{0.5cm}&\\
&~~\ld^{(2)}_{ku}={\sum_{\vec{v}}(-1)^{v_k-1}\prod_{k'\neq k}P_0\left(R_{v_{k'}|u}\right)\log{P_0\left(R_{{\vec{v}}|u}\right)\over P_1\left(R_{{\vec{v}}|u}\right)}\over \sum_{\vec{v}}(-1)^{v_k-1}\prod_{k'\neq k}P_1\left(R_{v_{k'}|u}\right){P_0\left(R_{{\vec{v}}|u}\right)\over P_1\left(R_{{\vec{v}}|u}\right)}},
\ea\ee
where $P_i\left(R_{{\vec{v}}|u}\right) = \prod_{k=1}^KP_i\left(R_{v_k|u}\right)$.
The degenerate decision regions of Lemma \ref{N-step-lemma} maintain their form as well. Since the decision rules have the same critical features (including threshold structure), our conclusions hold for this more general setup as well. This includes the multiple-step MIF of Section \ref{N-step process} with $K$ peripheral sensors, shown in Fig. \ref{fusion-diagrams-multiple-sensor2} for $N=3,5$ steps.

\subsection{Interactive fusion with soft sensor outputs}\label{multiple-bit}
We established in Section \ref{section_KL} that the two systems, namely the $YX$ and $XYX$ processes, have identical asymptotic detection performance when the sensor output is always binary. Consider the other extreme case where the exchange of information is endowed with unlimited bandwidth. In that case, entire observations can be exchanged between sensors and thus both the YX and XYX processes are tantamount to centralized detection. Therefore, the two systems again achieve exactly the same detection performance. It remains to see if that is still the case for interactive fusion when soft information is exchanged, i.e., sensor outputs are of multiple but finite number of bits.

Consider the case where $u$ and $v$ can take respectively $m$ and $l$ bits. Equivalently, we have $u\in \{0,1,...,2^m-1\}$ and $v\in \{0,1,...,2^l-1\}$. Improvement of performance in the fixed-sample NP test is immediate by induction since the single bit decisions are a particular case of the multiple bit decisions. Therefore we consider the situation for the asymptotic test.

By Proposition \ref{region_proposition}, the decision regions at $X$ are given by
\be\ba
\label{XYX-region_4} &R_{u=k}=\bigcap_{k'\neq k}\left\{x:{\del K[x,v]\over\del p(u=k|x)}-{\del K[x,v]\over\del p(u=k'|x)}>0\right\},\\
&~~~~k=0,1,...,2^m-1,
\ea\ee
and those at $Y$ are given by
\be\ba
\label{XYX-region_44}& R_{v=k|u}=\bigcap_{k'\neq k}\left\{y:{\del K[x,v]\over\del p(v=k|y,u)}-{\del K[x,v]\over\del p(v=k'|y,u)}>0\right\},\\
&~~~~k=0,1,...,2^l-1,
\ea\ee
where the objective function $K[x,v]$ is defined by (\ref{KL-XYX}). It is straightforward, with the help of equation (\ref{solution}), to verify that all the critical features of our analysis remain unchanged. In particular, by the same procedure as in the proofs of Theorem \ref{XYX_theorem}  and Lemma \ref{N-step-lemma}, the decision regions $R_{u=k}$ in (\ref{XYX-region_4}) have the form
\be
R_{u=k}=\bigcap_{k'\neq k}\left\{x:\sum_{k''=0}^{2^m-1}I_{R_{u=k''}}(x)a_{k''k'}>0\right\},~~k=0,1,...,2^m-1,
\ee
which admits a piecewise constant probability. Hence multiple bit passing before the final decision does not alter our results.

\begin{comment}
\subsection{The effect of randomization}
Recall the randomized decision rule~ $p_{\txt{opt}}(u=i|x)=I_{R_{u=i}}(x)+\sum_k\rho_{ik}I_{C_k}(x)$, where $\{C_k\}$ is a partition of the set $\bigcup_iC_{u=i}$ and $\rho_{ik}\in [0,1],~\sum_i\rho_{ik}=1$, and the sets $C_{u=i}$ are defined in (\ref{null_set}). As in the previous sections \ref{N-step process}, \ref{multiple-sensors}, and \ref{multiple-bit}, we can straightforwardly verify that our main results are not affected by randomization.
\end{comment}

\section{Conclusion}\label{concl}
We have considered a class of decision theory problems that involve convex objective functions and used it to study two-sensor tandem fusion networks with conditionally independent observations. We first established the optimum decision structure of a decision problem where a convex objective function is to be maximized. This result was then used to show that while interactive fusion improves the performance of fixed sample size NP test, it does not affect the asymptotic performance as characterized by the error exponent of type II error.

Several extensions of the above result were considered. The lack of improvement in the asymptotic detection performance of the one-step interactive fusion was shown to extend to multiple-step memoryless interactive fusion. Furthermore, the result was shown to be valid in a more general setting where the FC simultaneously interacted with $K\geq 1$ independent sensors as well as that involve multi-bit sensor output.

\appendices
\section{Proof of Theorem \ref{np-theorem}}\label{NP-test-details}
For conciseness, we embed the Lagrangian (\ref{NP-Lagrangian-0}) into a class of objective functions of the form\\
\bea
\label{uniform-objective}S=\sum_wf(p_0(w),p_1(w)),
\eea
where $f(x,y)$ is differentiable and convex in $(x,y)$, and hence differentiable and convex in any variables on which $x$ and $y$ depend linearly. Let $f_{iw}$ denote ${\del f(p_0(w),p_1(w))\over\del p_i(w)}$, which are, as we shall see, the essential \emph{threshold structure constants} that distinguish one objective function of the class (\ref{uniform-objective}) from another. In particular, for the Lagrangian (\ref{NP-Lagrangian-0}),
\bea
f_{01}=-\ld,~~f_{11}=1,~~f_{10}=0,~~f_{00}=0.
\eea

Using the dependence structure of the observation and decision variables, we can express the probability of the decision at the final step as in (\ref{NP-YX-probabilty}) and (\ref{NP-XYX-probabilty}). That is, for YX,
\bea
\label{NP-YX-probabiltyA}p_i(w)=\sum_{x,y,v}p(w|x,v)p(v|y)~p_i(x)p_i(y),
\eea
meanwhile for XYX,
\bea
\label{NP-XYX-probabiltyA}p_i(w)=\sum_{x,v,y,u}p(w|x,v)p(v|y,u)p(u|x)~p_i(x)p_i(y).
\eea

\subsection{The YX process}
By Proposition \ref{region_proposition} the decision regions for the YX process are the following.
For the decision $v$ at sensor Y,
\be\ba
&R_{v=1}=\left\{y:~{\del^B S\over\del p(v=1|y)}>0\right\}\sr{(a)}{=}\left\{y:~{p_1(y)\over p_0(y)}>\ld^{(2)}\right\},\nn\\
&\ld^{(2)}=\ld^{(0)}~{P_0(R_{w=1|v=1})-P_0(R_{w=1|v=0})\over P_1(R_{w=1|v=1})-P_1(R_{w=1|v=0})}\nn
\ea\ee
and for the decision $w$ at sensor X,
\be\ba
&R_{w=1|v}=\left\{x:{\del^B S\over\del p(w=1|x,v)}>0\right\}\sr{(b)}{=}\left\{x:{p_1(x)\over p_0(x)}>\ld_{v}^{(3)}\right\},\nn\\
&\ld_{v}^{(3)}=\ld^{(0)}~{P_0(R_{v})\over P_1(R_{v})},\nn
\ea\ee
where $\ld^{(0)}$ is as in (\ref{ld0}). Step (a) is given by
\be\ba
&{\del^B S\over\del p(v=1|y)}\sr{(\ref{uniform-objective})}{=}\sum_{w}{\del^B f(p_0(w),p_1(w))\over\del p(v=1|y)}=\sum_{i,w}f_{iw}{\del^B p_i(w)\over\del p(v=1|y)}\nn\\
&~~~~\sr{(\ref{NP-YX-probabiltyA})}{=} \sum_{i,w}f_{iw}\sum_{x,y',v'}p(w|x,v'){\del^B p(v'|y')\over\del p(v=1|y)}~p_i(x)p_i(y')\nn\\
&~~~~\sr{(\ref{binary-derivative})}{=} \sum_{i,w}f_{iw}\sum_{x,y',v}p(w|x,v)(-1)^{v-1}\delta_{yy'}~p_i(x)p_i(y')= \sum_{i,w}f_{iw}\sum_{x,v}p(w|x,v)(-1)^{v-1}~p_i(x)p_i(y)\nn\\
&~~~~\sr{(\ref{optimal1})}{=}\sum_{i,w}f_{iw}\sum_{x,v}I_{R_{w|v}}(x)(-1)^{v-1}~p_i(x)p_i(y)=\sum_{i,w}p_i(y)f_{iw}\sum_{v}P_i\left(R_{w|v}\right)(-1)^{v-1}\nn\\
&~~~~=\sum_{i,w}p_i(y)f_{iw}\left[P_i\left(R_{w|v=1}\right)-P_i\left(R_{w|v=0}\right)\right]\nn\\
&~~~~\sr{(a1)}{=}\sum_{i}p_i(y)(f_{i1}-f_{i0})\left[P_i\left(R_{w=1|v=1}\right)-P_i\left(R_{w=1|v=0}\right)\right]\nn\\
&~~~~=p_1(y)(f_{11}-f_{10})\left[P_1\left(R_{w=1|v=1}\right)-P_1\left(R_{w=1|v=0}\right)\right]\nn\\
&~~~~~~~~+p_0(y)(f_{01}-f_{00})\left[P_0\left(R_{w=1|v=1}\right)-P_0\left(R_{w=1|v=0}\right)\right]\nn\\
&~~~~=p_0(y)(f_{11}-f_{10})\left[P_1\left(R_{w=1|v=1}\right)-P_1\left(R_{w=1|v=0}\right)\right]\nn\\
&~~~~~~~~\times \left[{p_1(y)\over p_0(y)}-{f_{00}-f_{01}\over f_{11}-f_{10}}~{P_0\left(R_{w=1|v=1}\right)-P_0\left(R_{w=1|v=0}\right)\over P_1\left(R_{w=1|v=1}\right)-P_1\left(R_{w=1|v=0}\right)}\right]\nn\\
&~~~~=p_0(y)(f_{11}-f_{10})\left[P_1\left(R_{w=1|v=1}\right)-P_1\left(R_{w=1|v=0}\right)\right]\left[{p_1(y)\over p_0(y)}-\ld^{(2)}\right],\nn
\ea\ee
where step $(a1)$ uses $P_i(R_{w=0|v})=1-P_i(R_{w=0|v})$, and the rest of the labeled steps $\sr{(...)}{=}$ reference equations from which the equality comes.

Likewise, step (b) is given by
\be\ba
&{\del^B S\over\del p(w=1|x,v)}\sr{(\ref{uniform-objective})}{=}\sum_{w}{\del^B f(p_0(w),p_1(w))\over\del p(w=1|x,v)}=\sum_{i,w}f_{iw}{\del^B p_i(w)\over\del p(w=1|x,v)}\nn\\
&~~~~\sr{(\ref{NP-YX-probabiltyA})}{=} \sum_{i,w'}f_{iw'}\sum_{x',y,v'}{\del^B p(w'|x',v')\over\del p(w=1|x,v)}p(v'|y)~p_i(x')p_i(y)\nn\\
&~~~~\sr{(\ref{binary-derivative})}{=} \sum_{i,w}f_{iw}\sum_{x',y,v'}(-1)^{w-1}\delta_{xx'}\delta_{vv'}p(v'|y)~p_i(x')p_i(y)= \sum_{i,w}f_{iw}\sum_{y}(-1)^{w-1}p(v|y)~p_i(x)p_i(y)\nn\\
&~~~~\sr{(\ref{optimal1})}{=}\sum_{i,w}f_{iw}\sum_{y}(-1)^{w-1}I_{R_v}(y)~p_i(x)p_i(y)=\sum_{i,w}p_i(x)f_{iw}(-1)^{w-1}P_i\left(R_v\right)=\sum_{i}p_i(x)(f_{i1}-f_{i0})P_i\left(R_v\right)\nn\\
&~~~~=p_1(x)(f_{11}-f_{10})P_1\left(R_v\right)+p_0(x)(f_{01}-f_{00})P_0\left(R_v\right)\nn\\
&~~~~=p_0(x)(f_{11}-f_{10})P_1\left(R_v\right)\left[{p_1(x)\over p_0(x)}-{f_{00}-f_{01}\over f_{11}-f_{10}}~{P_0\left(R_v\right)\over P_1\left(R_v\right)}\right]\nn\\
&~~~~=p_0(x)(f_{11}-f_{10})P_1\left(R_v\right)\left[{p_1(x)\over p_0(x)}-\ld^{(3)}_v\right],\nn
\ea\ee
where the labeled steps $\sr{(...)}{=}$ reference equations from which the equality comes.

With { $p_{i\max}(w)= \sum_vP_i(R_{w|v})P_i(R_v)$}, the maximum value of $S$ is
\be
S_{\max}= \sum_wf\left(\sum_{v}P_0(R_{v})P_0(R_{w|v})~,\sum_{v}P_1(R_{v})P_1(R_{w|v})\right).\nn
\ee

\subsection{The XYX process}
Similarly, for the XYX process, we obtain the following decision regions.
For the first decision $u$ at sensor X,
\be
\label{first-xyx-region-apdx}R_{u=1}=\left\{x:~{\del^B S\over\del p(u=1|x)}>0\right\}\sr{(c)}{=}\left\{x:~{p_1(x)\over p_0(x)}Q(x)>\ld^{(1)}Q(x)\right\},
\ee
where { $\ld^{(1)}=\ld^{(0)}~{P_0(R_{v=1|u=1})-P_0(R_{v=1|u=0})\over P_1(R_{v=1|u=1})-P_1(R_{v=1|u=0})},$}
\bea
\label{ld0}\ld^{(0)}={{\del f(p_0(w),p_1(w))\over \del p_0(w)}\big|_{w=0}-{\del f(p_0(w),p_1(w))\over \del p_0(w)}\big|_{w=1}\over {\del f(p_0(w),p_1(w))\over \del p_1(w)}\big|_{w=1}-{\del f(p_0(w),p_1(w))\over \del p_1(w)}\big|_{w=0}},
\eea
and  {$Q(x)=I_{R_{w=1|v=1}}(x)-I_{R_{w=1|v=0}}(x)$}. Step (c) involves the following evaluations.
\be\ba
&{\del^B S\over\del p(u=1|x)}\sr{(\ref{uniform-objective})}{=}\sum_{w}{\del^B f(p_0(w),p_1(w))\over\del p(u=1|x)}=\sum_{i,w}f_{iw}{\del^B p_i(w)\over\del p(u=1|x)}\nn\\
&~~~~~\sr{(\ref{NP-XYX-probabiltyA})}{=}~\sum_{i,w}f_{iw}\sum_{x',v,y,u'}p(w|x',v)p(v|y,u'){\del^B p(u'|x')\over\del p(u=1|x)}~p_i(x')p_i(y)\nn\\
&~~~~\sr{(\ref{binary-derivative})}{=}\sum_{i,w}f_{iw}\sum_{x',v,y,u}p(w|x',v)p(v|y,u)(-1)^{u-1}\delta_{xx'}~p_i(x')p_i(y)\nn\\
&~~~~=\sum_{i,w}f_{iw}\sum_{v,y,u}p(w|x,v)p(v|y,u)(-1)^{u-1}~p_i(x)p_i(y)\nn\\
\ea\ee
\be\ba
&~~~~\sr{(\ref{optimal1})}{=}\sum_{i,w}f_{iw}\sum_{v,y,u}I_{R_{w|v}}(x)I_{R_{v|u}}(y)(-1)^{u-1}~p_i(x)p_i(y)\nn\\
&~~~~=\sum_{i,w}p_i(x)f_{iw}\sum_{v}I_{R_{w|v}}(x)\sum_uP_i(R_{v|u})(-1)^{u-1}\nn\\
&~~~~=\sum_{i,w}p_i(x)f_{iw}\sum_{v}I_{R_{w|v}}(x)\left(P_i(R_{v|u=1})-P_i(R_{v|u=0})\right)\nn\\
&~~~~\sr{(c1)}{=}\sum_{i,w}p_i(x)f_{iw}\left(I_{R_{w|v=1}}(x)-I_{R_{w|v=0}}(x)\right)\left(P_i(R_{v=1|u=1})-P_i(R_{v=1|u=0})\right)\nn\\
&~~~~\sr{(c2)}{=}\sum_{i}p_i(x)(f_{i1}-f_{i0})\left(I_{R_{w=1|v=1}}(x)-I_{R_{w=1|v=0}}(x)\right)\left(P_i(R_{v=1|u=1})-P_i(R_{v=1|u=0})\right)\nn\\
&~~~~=\left(I_{R_{w=1|v=1}}(x)-I_{R_{w=1|v=0}}(x)\right)\sum_{i}p_i(x)(f_{i1}-f_{i0})\left(P_i(R_{v=1|u=1})-P_i(R_{v=1|u=0})\right)\nn\\
&~~~~=Q(x)\left[p_1(x)(f_{11}-f_{10})\left(P_1(R_{v=1|u=1})-P_1(R_{v=1|u=0})\right)\right.\nn\\
&~~~~~~~~\left.+p_0(x)(f_{01}-f_{00})\left(P_0(R_{v=1|u=1})-P_0(R_{v=1|u=0})\right)\right]\nn\\
&~~~~=p_0(x)(f_{11}-f_{10})\left(P_1(R_{v=1|u=1})-P_1(R_{v=1|u=0})\right)\nn\\
&~~~~~~~~\times Q(x)\left[{p_1(x)\over p_0(x)}-{f_{00}-f_{01}\over f_{11}-f_{10}}~{P_0(R_{v=1|u=1})-P_0(R_{v=1|u=0})\over P_1(R_{v=1|u=1})-P_1(R_{v=1|u=0})}\right]\nn\\
&~~~~=p_0(x)(f_{11}-f_{10})\left(P_1(R_{v=1|u=1})-P_1(R_{v=1|u=0})\right)Q(x)\left[{p_1(x)\over p_0(x)}-\ld^{(1)}\right],\nn
\ea\ee
where step $(c1)$ uses $P_i(R_{v=0|u})=1-P_i(R_{v=0|u})$, step $(c2)$ uses $I_{R_{w=0|v}}(x)=I_{\X}(x)-I_{R_{w=1|v}}(x)$, and the rest of the labeled steps $\sr{(...)}{=}$ reference equations from which the equality comes.

For the decision $v$ at sensor Y,
\be\ba
\label{second-xyx-region-apdx} R_{v=1|u}=\left\{y:~{\del^B S\over\del p(v=1|y,u)}>0\right\}\sr{(d)}{=}\left\{y:~{p_1(y)\over p_0(y)}>\ld^{(2)}_u\right\},
\ea\ee
where { $\ld^{(2)}_u=\ld^{(0)}~{P_0(R_{w=1|v=1}\cap R_u)-P_0(R_{w=1|v=0}\cap R_u)\over P_1(R_{w=1|v=1}\cap R_u)-P_1(R_{w=1|v=0}\cap R_u)}$}. Step $(d)$ is derived as follows.
\be\ba
&{\del^B S\over\del p(v=1|y,u)}\sr{(\ref{uniform-objective})}{=}\sum_{w}{\del^B f(p_0(w),p_1(w))\over\del p(v=1|y,u)}=\sum_{i,w}f_{iw}{\del^B p_i(w)\over\del p(v=1|y,u)}\nn\\
&~~~\sr{(\ref{NP-XYX-probabiltyA})}{=}\sum_{i,w}f_{iw}\sum_{x,v',y',u'}p(w|x,v'){\del^B p(v'|y',u')\over\del p(v=1|y,u)}~p(u'|x)p_i(x)p_i(y')\nn\\
&~~~\sr{(\ref{binary-derivative})}{=}\sum_{i,w}f_{iw}\sum_{x,v,y',u'}p(w|x,v)(-1)^{v-1}\delta_{yy'}\delta_{uu'}~p(u'|x)p_i(x)p_i(y')\nn\\
&~~~=\sum_{i,w}f_{iw}\sum_{x,v}p(w|x,v)(-1)^{v-1}~p(u|x)p_i(x)p_i(y)\nn\\
&~~~\sr{(\ref{optimal1})}{=}\sum_{i,w}f_{iw}\sum_{x,v}I_{R_{w|v}}(x)(-1)^{v-1}I_{R_u}(x)p_i(x)p_i(y)\nn\\
&~~~=\sum_{i,w}f_{iw}\sum_{x,v}I_{R_{w|v}\cap R_u}(x)(-1)^{v-1}p_i(x)p_i(y)=\sum_{i,w}p_i(y)f_{iw}\sum_{v}P_i\left(R_{w|v}\cap R_u\right)(-1)^{v-1}\nn\\
&~~~=\sum_{i,w}p_i(y)f_{iw}\left[P_i\left(R_{w|v=1}\cap R_u\right)-P_i\left(R_{w|v=0}\cap R_u\right)\right]\nn\\
\ea\ee
\be\ba
&~~~\sr{(d1)}{=}\sum_{i}p_i(y)(f_{i1}-f_{i0})\left[P_i\left(R_{w=1|v=1}\cap R_u\right)-P_i\left(R_{w=1|v=0}\cap R_u\right)\right]\nn\\
&~~~=p_1(y)(f_{11}-f_{10})\left[P_1\left(R_{w=1|v=1}\cap R_u\right)-P_1\left(R_{w=1|v=0}\cap R_u\right)\right]\nn\\
&~~~~+p_0(y)(f_{01}-f_{00})\left[P_0\left(R_{w=1|v=1}\cap R_u\right)-P_0\left(R_{w=1|v=0}\cap R_u\right)\right]\nn\\
&~~~=p_0(y)(f_{11}-f_{10})\left[P_1\left(R_{w=1|v=1}\cap R_u\right)-P_1\left(R_{w=1|v=0}\cap R_u\right)\right]\nn\\
&~~~~\times\left[{p_1(y)\over p_0(y)}-\ld^{(0)}{P_0\left(R_{w=1|v=1}\cap R_u\right)-P_0\left(R_{w=1|v=0}\cap R_u\right)\over P_1\left(R_{w=1|v=1}\cap R_u\right)-P_1\left(R_{w=1|v=0}\cap R_u\right)}\right]\nn\\
&~~~=p_0(y)(f_{11}-f_{10})\left[P_1\left(R_{w=1|v=1}\cap R_u\right)-P_1\left(R_{w=1|v=0}\cap R_u\right)\right]\left[{p_1(y)\over p_0(y)}-\ld^{(2)}_u\right],\nn
\ea\ee
where step $(d1)$ uses $P_i\left(R_{w=0|v}\cap R_u\right)=P_i\left(R_u\right)-P_i\left(R_{w=1|v}\cap R_u\right)$, and the rest of the labeled steps $\sr{(...)}{=}$ reference equations from which the equality comes.

Finally, for the second decision $w$ at sensor X,
\be\ba
\label{third-xyx-region-apdx} R_{w=1|v}\!=\!\left\{x:{\del^B S\over\del p(w=1|x,v)}>0\right\}\!\sr{(e)}{=}\!\left\{x:{p_1(x)\over p_0(x)}\!>\!\sum_u\ld_{vu}^{(3)}I_{R_u}(x)\right\},
\ea\ee
where { $\ld_{vu}^{(3)}=\ld^{(0)}~{P_0(R_{v|u})\over P_1(R_{v|u})}$}. Note that only two out of these four thresholds are independent. Step $(e)$ is derived as follows.
\be\ba
&{\del^B S\over\del p(w=1|x,v)}\sr{(\ref{uniform-objective})}{=}\sum_{w}{\del^B f(p_0(w),p_1(w))\over\del p(w=1|x,v)}=\sum_{i,w}f_{iw}{\del^B p_i(w)\over\del p(w=1|x,v)}\nn\\
&~~~\sr{(\ref{NP-XYX-probabiltyA})}{=}~\sum_{i,w'}f_{iw'}\sum_{x',v',y,u}{\del^B p(w'|x',v')\over\del p(w=1|x,v)}p(v'|y,u)~p(u|x')p_i(x')p_i(y)\nn\\
&~~\sr{(\ref{binary-derivative})}{=}~\sum_{i,w}f_{iw}\sum_{x',v',y,u}(-1)^{w-1}\delta_{xx'}\delta_{vv'}p(v'|y,u)~p(u|x')p_i(x')p_i(y)\nn\\
&~~=~\sum_{i,w}f_{iw}\sum_{y,u}(-1)^{w-1}p(v|y,u)~p(u|x)p_i(x)p_i(y)\nn\\
&~~\sr{(\ref{optimal1})}{=}~\sum_{i,w}f_{iw}(-1)^{w-1}\sum_{y,u}I_{R_{v|u}}(y)I_{R_u}(x)p_i(x)p_i(y)=\sum_{i}p_i(x)(f_{i1}-f_{i0})\sum_{u}P_i\left(R_{v|u}\right)I_{R_u}(x)\nn\\
&~~=~p_0(x)(f_{11}-f_{10})\sum_{u}P_1\left(R_{v|u}\right)I_{R_u}(x)+p_0(x)(f_{01}-f_{00})\sum_{u}P_0\left(R_{v|u}\right)I_{R_u}(x)\nn\\
&~~=~p_0(x)(f_{11}-f_{10})\sum_{u}P_1\left(R_{v|u}\right)I_{R_u}(x)\left[{p_1(x)\over p_0(x)}-{f_{00}-f_{01}\over f_{11}-f_{10}}~{\sum_{u}P_0\left(R_{v|u}\right)I_{R_u}(x)\over \sum_{u}P_1\left(R_{v|u}\right)I_{R_u}(x)}\right]\nn\\
&~~\sr{(\ref{identity1})}{=}~p_0(x)(f_{11}-f_{10})\sum_{u}P_1\left(R_{v|u}\right)I_{R_u}(x)\left[{p_1(x)\over p_0(x)}-{f_{00}-f_{01}\over f_{11}-f_{10}}~\sum_u{P_0\left(R_{v|u}\right)\over P_1\left(R_{v|u}\right)}I_{R_u}(x)\right]\nn\\
&~~=p_0(x)(f_{11}-f_{10})\sum_{u}P_1\left(R_{v|u}\right)I_{R_u}(x)\left[{p_1(x)\over p_0(x)}-\sum_u\ld^{(3)}_{vu}I_{R_u}(x)\right],\nn
\ea\ee
where the labeled steps $\sr{(...)}{=}$ reference equations from which the equality comes.

The probability of the final decision at the optimal point is { $p_{i\max}(w)= \sum_{u,v}P_i(R_{v|u})P_i(R_{w|v}\cap R_u)$}, and the maximum value $S_{\max}$ of the objective function is
\be
\sum_wf\left(\sum_{u,v}P_0(R_{v|u})P_0(R_{w|v}\cap R_u)~,\sum_{u,v}P_1(R_{v|u})P_1(R_{w|v}\cap R_u)\right).\nn
\ee
The region $R_{w=1|v}$ in (\ref{third-xyx-region-apdx}) can be written as~ { $R_{w=1|v}=\bigcup_uR_{w=1|v,u},$} where\\ { $R_{w=1|v,u}=\left\{x:~p_1(x)/p_0(x)>\ld_{vu}^{(3)}\right\}$} and we recall that $\ld_{vu}^{(3)}$ has only two independent components. For our case of interest where the objective  $S$ is the Lagrangian $L$ in equation (\ref{NP-Lagrangian-0}),~ we can easily check that ~ $\ld^{(0)}=\ld,$ with
\be
\label{fixed-YX-lagrangian}L=\sum_{v}P_1(R_{v})P_1(R_{w=1|v}) +\ld~\left[\al-\sum_{v}P_0(R_{v})P_0(R_{w=1|v})\right]
\ee
for the YX Process, and $L$ is
\be
\label{fixed-XYX-lagrangian}\sum_{u,v}P_1(R_{v|u})P_1(R_{w=1|v}\cap R_u)+\ld~\left[\al-\sum_{u,v}P_0(R_{v|u})P_0(R_{w=1|v}\cap R_u)\right]
\ee
for the XYX Process.

\section{Details of the fixed sample size NP test for a constant signal in WGN}\label{wgn-details}
Consider the example (\ref{WGN}) for a constant signal in white Gaussian noise. The YX regions and their probabilities are as follows.

\be\ba
& R_{v=1}=\left\{y>t^{(2)}=\sigma^2_y\log\ld^{(2)}+1/2\right\},\nn\\
& R_{w=1|v}=\left\{x>t^{(3)}_v=\sigma^2\log\ld^{(3)}_v+1/2\right\},\nn\\
&P_i(R_{v})=\left[Q\left({t^{(2)}-i\over\sigma_y}\right)\right]^v\left[1-Q\left({t^{(2)}-i\over\sigma_y}\right)\right]^{1-v},\nn\\
&P_i(R_{w|v})=\left[Q\left({t^{(3)}_{v}-i\over\sigma_x}\right)\right]^w\left[1-Q\left({t^{(3)}_{v}-i\over\sigma_x}\right)\right]^{1-w},\nn
\ea\ee

where $\ld^{(2)}$ and $\ld^{(3)}_v$ are defined as in (\ref{NP-YX-regions}).

Observe that the XYX decision rule is invariant if we switch the two decision regions for $w$ and the two decision regions for $v$, i.e.,
\bea
\label{exchange-trans}R_{w=1|v=1}\longleftrightarrow R_{w=1|v=0},~~~~R_{v=1|u=1} \longleftrightarrow R_{v=1|u=0}.
\eea
In addition, our Gaussian example has a monotone likelihood ratio (i.e., ${p_1(z)/p_0(z)}$ is a monotone function of $z$). Therefore, without loss of generality, we assume that $R_{v=1|u=0}\subset R_{v=1|u=1}$, ~ $R_{w=1|v=0}\subset R_{w=1|v=1}$. This assumption implies that { $Q(x)=I_{R_{w=1|v=1}}(x)-I_{R_{w=1|v=0}}(x)>0$,~ $P_i(R_{v=1|u=1})-P_i(R_{v=1|u=0})>0$}, which simplifies the region $R_{u=1}$ to a region defined by an LRT. Thus with $\ld^{(1)}$, $\ld^{(2)}_u$, and $\ld^{(3)}_{vu}$ defined as in (\ref{NP-XYX-regions}), the decision regions are given by
\begin{equation}
   \begin{aligned}
\label{XYX-simple}& R_{u=1}=\left\{x>t^{(1)}=\sigma^2_x\log\ld^{(1)}+1/2\right\},\\
& R_{v=1|u}=\left\{y>t^{(2)}_u=\sigma_y^2\log\ld^{(2)}_u+1/2\right\},\\
& R_{w=1|v}=\bigcup_uR_{w=1|v,u},\\
& R_{w=1|v,u}=\left\{x>t^{(3)}_{vu}=\sigma^2_x\log\ld^{(3)}_{vu}+1/2\right\},
   \end{aligned}
\end{equation}
and their probabilities are
\be\ba
& P_i(R_{u})=\left[Q\left({t^{(1)}-i\over\sigma_x}\right)\right]^u\left[1-Q\left({t^{(1)}-i\over\sigma_x}\right)\right]^{1-u},\nn\\
&P_i(R_{v|u})=\left[Q\left({t^{(2)}_u-i\over\sigma_y}\right)\right]^v\left[1-Q\left({t^{(2)}_u-i\over\sigma_y}\right)\right]^{1-v},\nn\\
&P_i(R_{w|v,u})=\left[Q\left({t^{(3)}_{vu}-i\over\sigma_x}\right)\right]^w\left[1-Q\left({t^{(3)}_{vu}-i\over\sigma_x}\right)\right]^{1-w},\nn\\
&P_i(R_{w|v})=P_i(R_{w|v,u=0}\cup R_{w|v,u=1})= P_i(R_{w|v,u=0})+P_i(R_{w|v,u=1})-P_i(R_{w|v,u=0}\cap R_{w|v,u=1}),\nn\\
&P_i(R_{w|v}\cap R_u)=P_i\big([R_{w|v,u=0}\cup R_{w|v,u=1}]\cap R_u\big)\nn\\
&~~~~= P_i(R_{w|v,u=0}\cap R_u)+P_i(R_{w|v,u=1}\cap R_u)-P_i(R_{w|v,u=0}\cap R_{w|v,u=1}\cap R_u),\nn
\ea\ee
where
\be\ba
  &P_i(R_{w=1|v=0}\cap R_{u=0})=Q\left({t^{(3)}_{00}-i\over \sigma_x}\right)-Q\left({\max(t^{(3)}_{00},t^{(1)})-i\over \sigma_x}\right)\nn\\
  &P_i(R_{w=1|v=0}\cap R_{u=1})=Q\left({\max(t^{(3)}_{00},t^{(1)})-i\over \sigma_x}\right)\nn\\
  &P_i(R_{w=1|v=1}\cap R_{u=0})=Q\left({t^{(3)}_{10}-i\over \sigma_x}\right)-Q\left({\max(t^{(3)}_{10},t^{(1)})-i\over \sigma_x}\right)\nn\\
  &P_i(R_{w=1|v=1}\cap R_{u=1})=Q\left({\max(t^{(3)}_{10},t^{(1)})-i\over \sigma_x}\right)\nn\\
  &P_i(R_{w=0|v=0}\cap R_{u=0})=1-Q\left({t^{(1)}-i\over\sigma_x}\right)-Q\left({t^{(3)}_{00}-i\over \sigma_x}\right)+Q\left({\max(t^{(3)}_{00},t^{(1)})-i\over \sigma_x}\right)\nn \\
  &P_i(R_{w=0|v=0}\cap R_{u=1})=Q\left({t^{(1)}-i\over\sigma_x}\right)-Q\left({\max(t^{(3)}_{00},t^{(1)})-i\over \sigma_x}\right)\nn\\
  &P_i(R_{w=0|v=1}\cap R_{u=0})=1-Q\left({t^{(1)}-i\over\sigma_x}\right)-Q\left({t^{(3)}_{10}-i\over \sigma_x}\right)+Q\left({\max(t^{(3)}_{10},t^{(1)})-i\over \sigma_x}\right)\nn \\
  &P_i(R_{w=0|v=1}\cap R_{u=1})=Q\left({t^{(1)}-i\over\sigma_x}\right)-Q\left({\max(t^{(3)}_{10},t^{(1)})-i\over \sigma_x}\right)\nn
\ea\ee
Notice from (\ref{XYX-simple}) and (\ref{NP-XYX-regions}) that the thresholds are coupled by relations of the form
\be\ba
\label{XYX-threshold-coupling}& t^{(1)}=g^{(1)}\big(t^{(1)},t^{(2)}_{u},t^{(3)}_{vu}\big),\\
& t^{(2)}_u=g^{(2)}_u\big(t^{(1)},t^{(2)}_{u},t^{(3)}_{vu}\big),\\
& t^{(3)}_{vu}=g^{(3)}_{vu}\big(t^{(1)},t^{(2)}_{u},t^{(3)}_{vu}\big),
\ea\ee
where $g^{(1)},~~g^{(2)}_u,~~g^{(3)}_{uv}$~ (with $g^{(3)}_{uv}$ having only two independent components) are determined by dependence of the probabilities of the decision regions on the thresholds. Numerical iteration can thus be used to compute the thresholds. Note however that by Remark 2 following Proposition \ref{region_proposition}, we can alternatively substitute the decision regions (\ref{XYX-simple}) into the objective function, i.e., the Lagrangian, (\ref{fixed-XYX-lagrangian}) and then directly optimize the Lagrangian over the thresholds.

Due to the simplifying assumptions that follow upon noticing invariance of the XYX decision rule under the transformation (\ref{exchange-trans}), initial conditions, i.e., initial values for the thresholds in (\ref{XYX-simple}) during numerical computation via iteration must be chosen to satisfy $t^{(3)}_{1u}<t^{(3)}_{0u}$,~ $t^{(3)}_{v0}<t^{(3)}_{v1}$, and $t^{(2)}_1<t^{(2)}_0,$ meanwhile $t^{(1)}$ can take any initial value.

If properly implemented, the above details lead to the graphs corresponding to the YX and XYX processes in Figure \ref{np-graph}.

\section{Proof of Lemma \ref{N-step-lemma}}\label{proof-N-step-lemma}
 Observe that the KL distance (\ref{N-step-KL-1}) has the form
\bea
\label{N-step-KL-2} K[x,u_{N-1}]=K[x]+\sum_{x}p_0(x)K[u_{N-1}|x],
\eea
where~ {\small $K[u_{N-1}|x]=\sum_{u_{N-1}}p_0(u_{N-1}|x)\log{p_0(u_{N-1}|x)\over p_1(u_{N-1}|x)}$}~
is independent of $p_0(x)$ and $p_1(x)$, since from the even $N$ case of (\ref{odd-case2}),
\be\ba \label{p-max1}  &p_i(u_{N-1}|x)=\sum_{y,u^{N-2}}p_i(y)\prod_{r=1}^{(N-1)/2}\left[p(u_{2r-1}|x,u_{2r-2})p(u_{2r}|y,u_{2r-1})\right]\sr{(a)}{=}\sum_{u^{N-2}}I_{\R_{u^{N-1}}}(x)~\rho_{i,u^{N-1}},\\
&\R_{u^{N-1}}=\bigcap_{r=1}^{(N-1)/2}R_{u_{2r-1}|u_{2r-2}},~~~~\rho_{i,u^{N-1}}=\sum_yp_i(y) \prod_{r=1}^{(N-1)/2}p(u_{2r}|y,u_{2r-1})\sr{(b)}{=}~P_i\left(\bigcap_{r=1}^{(N-1)/2}R_{u_{2r}|u_{2r-1}}\right),
\ea\ee
where equalities (a) and (b) are due to (\ref{optimal1}).
By Proposition \ref{region_proposition} we have that
\be
R_{u_{2r-1}=1|u_{2r-2}}=\left\{x:~ {\del^B K[x,u_{N-1}]\over\del p_{\max}(u_{2r-1}=1|x,u_{2r-2})}>0\right\}.
\ee
To evaluate {\small ${\del^B K[x,u_{N-1}]\over\del p_{\max}(u_{2r-1}=1|x,u_{2r-2})}$}, we first evaluate {\footnotesize ${\del^B p_i(u'_{N-1}|x')\over\del p(u_{2r-1}=1|x,u_{2r-2})}$} as follows.
%\begin{comment}
\bea
&&{\del^B p_i(u'_{N-1}|x')\over\del p(u_{2r-1}=1|x,u_{2r-2})}=\sum_{y,u'{}^{N-2}}p_i(y){\del^B p(u'_{2r-1}|x',u'_{2r-2})\over\del p(u_{2r-1}=1|x,u_{2r-2})}\nn\\
&&~~~~~~~~\times p(u'_{2r}|y,u'_{2r-1})\prod_{s\neq r}^{(N-1)/2}p(u'_{2s-1}|x',u'_{2s-2})p(u'_{2s}|y,u'_{2s-1})\nn\\
&&~~~~=\delta_{xx'}\sum_{y,u'{}^{N-2}}p_i(y) (-1)^{u'_{2r-1}-1}\delta_{u'_{2r-2}u_{2r-2}}p(u'_{2r}|y,u'_{2r-1})\prod_{s\neq r}^{(N-1)/2}p(u'_{2s-1}|x',u'_{2s-2})p(u'_{2s}|y,u'_{2s-1})\nn\\
\label{p-derivative1}&&~~~~~\sr{(a)}{=}\delta_{xx'}\sum_{u'{}^{N-2}}I_{\R_{u_r'{}^{N-1}}}(x)~\rho_{i,u'{}^{N-1}|u_{2r-2}},\\
&&\R_{u_r'{}^{N-1}} = \bigcap_{s\neq r}^{(N-1)/2}R_{u'_{2s-1}|u'_{2s-2}},\nn\\
&&\rho_{i,u'{}^{N-1}|u_{2r-2}} =\sum_yp_i(y) (-1)^{u'_{2r-1}-1}\delta_{u'_{2r-2}u_{2r-2}}\prod_{s=1}^{(N-1)/2}p(u'_{2s}|y,u'_{2s-1})\nn
\eea
\bea
&&~~~~\sr{(b)}{=}(-1)^{u'_{2r-1}-1}\delta_{u'_{2r-2}u_{2r-2}} P_i\left(\bigcap_{s=1}^{(N-1)/2}R_{u'_{2s}|u'_{2s-1}}\right),\nn
\eea
where equalities $(a)$ and $(b)$ come from (\ref{optimal1}).
Therefore from (\ref{N-step-KL-2}),
%\end{comment}
\be\ba
& {\del^B K[x,u_{N-1}]\over\del p_{\max}(u_{2r-1}=1|x,u_{2r-2})}=\sum_{x'}p_0(x'){\del^B K[u_{N-1}|x']\over\del p_{\max}(u_{2r-1}=1|x,u_{2r-2})}\nn\\
&~~~~=\sum_{x',u'_{N-1}}p_0(x')\left[{\del^B p_0(u'_{N-1}|x)\over\del p_{\max}(u_{2r-1}=1|x,u_{2r-2})}\left(\log{p_0(u'_{N-1}|x')\over p_1(u'_{N-1}|x')}+1\right)\right.\nn\\
&~~~~~~~~\left.-{p_0(u'_{N-1}|x')\over p_1(u'_{N-1}|x')}{\del^B p_1(u'_{N-1}|x')\over\del p_{\max}(u_{2r-1}=1|x,u_{2r-2})} \right]\nn\\
&~~~~\sr{(a)}{=}\sum_{x,x',u'{}^{N-1}}p_0(x')\delta_{xx'}I_{\R_{u_r'{}^{N-1}}}(x)\left[\rho_{0,u'{}^{N-1}|u_{2r-2}}\right.\nn\\
&~~~~~~~~\left.\times\left(\log{p_0(u'_{N-1}|x')\over p_1(u'_{N-1}|x')}+1\right)-{p_0(u'_{N-1}|x')\over p_1(u'_{N-1}|x')}\rho_{1,u'{}^{N-1}|u_{2r-2}} \right]\nn\\
&~~~~\sr{(b)}{=}p_0(x)\sum_{u'{}^{N-1}}I_{\R_{u_r'{}^{N-1}}}(x)\left[\log{p_0(u'_{N-1}|x)\over p_1(u'_{N-1}|x)}~\rho_{0,u'{}^{N-1}|u_{2r-2}}-{p_0(u'_{N-1}|x)\over p_1(u'_{N-1}|x)}~\rho_{1,u'{}^{N-1}|u_{2r-2}} \right]\nn\\
&~~~~~\sr{(c)}{=}~p_0(x)\sum_{\al}I_{D_{\al}}(x) A_{\al,r,u_{2r-2}},
\begin{comment}
&&~~~~=p_0(x)\sum_{u'{}^{N-1},u''^{N-2}}I_{D_{u_r'{}^{N-1}}}(x)I_{D_{(u'_{N-1},u''{}^{N-2})}}(x)\left[\rho_{0,u'{}^{N-1}|u_{2r-2}}\left(\log{\rho_{0,(u'_{N-1},u''{}^{N-2})}\over \rho_{1,(u'_{N-1},u''{}^{N-2})}}+1\right)-\log{\rho_{0,(u'_{N-1},u''{}^{N-2})}\over \rho_{1,(u'_{N-1},u''{}^{N-2})}}\rho_{1,u'{}^{N-1}|u_{2r-2}} \right]\nn\\
\end{comment}
\ea\ee
where (a) is from equation (\ref{p-derivative1}), (c) from equations (\ref{identity1}) and (\ref{p-max1}), $\{D_\al\}$ is some partition of the data space $\X$, and the coefficients $A_{\al,r,u_{2r-2}}$ do not depend on $x$. The second term of the previous step evaluates to zero at step $(b)$. Note that from step $(b)$, in order to apply (\ref{identity1}) at the next step, we have chosen a partition that allows us to first write for each $r\in\{1,...,(N-1)/2\}$,
\be
I_{R_{u_{2r-1}=1|u_{2r-2}}}(x) = \sum_{\al_r\in\{\al_r\}}I_{D_{\al_r}}(x),~~\al_r=\al_r(u_{2r-2}),\nn
\ee
for some subset $\{\al_r\}\subset \{\al\}$ of the index set $\{\al\}$.

%%%%%%%%%%%%%%%%%%%%%%%%%%%%%%%%%%%%%%%%%%%%%%%%%%%%%%%%%%%%%%%%%%%%%%%%%%%%%%%%%%%%%%%%%%%%%%%%%%%%%%%%%%%%%%%%%%%%%


\end{document}